\documentclass[aps, twocolumn,floatfix,superscriptaddress,longbibliography]{revtex4-2}
\usepackage[utf8]{inputenc}
\usepackage{natbib}
\usepackage[normalem]{ulem}
\usepackage{graphicx}
\usepackage{xcolor}

\usepackage[tbtags]{amsmath}
\usepackage{hyperref}
\hypersetup{colorlinks,allcolors=black}
\usepackage{amssymb}
\usepackage{gensymb}
\usepackage{float}
\usepackage{amsmath}
\usepackage{tabularx,graphicx}
\usepackage{epstopdf}
\usepackage{latexsym}
\usepackage{color, colortbl}
\usepackage{psfrag}
\usepackage{bbm}
\usepackage{bm}
\usepackage{titlesec}
\usepackage{dsfont}
\usepackage{feynmp}
\usepackage{slashed}
\usepackage{multirow}
\usepackage{soul} 
\usepackage{braket}
\usepackage{caption}
\usepackage{subcaption}
\usepackage{pgfplots}
\usepackage{siunitx}
\usepackage[Export]{adjustbox} 
\adjustboxset{max size={0.9\linewidth}{0.9\paperheight}}
 \usepackage{commath}
 \usepackage{mathtools}
\usepackage{listings}
 \usepackage{comment}
 \usepackage{wrapfig}
 \usepackage{subfiles}
 \usepackage{multirow}
\usetikzlibrary{plotmarks}

\newcommand{\tr}{{\rm tr\,}}

\renewcommand{\hat}[1]{{\bf {\widehat #1}}}
\renewcommand{\phi}{\varphi}
\renewcommand{\epsilon}{\varepsilon}

\let\vaccent=\v 
\renewcommand{\v}[1]{\ensuremath{\mathbf{#1}}} 
\newcommand{\half}{\frac{1}{2}}

\newcommand{\dtop}[1]{d_{#1}^{\text{top}}}
\newcommand{\dbot}[1]{d_{#1}^{\text{bottom}}}
\newcommand{\thetaeff}[1]{\theta^{\text{eff}}_{#1}}
\newcommand{\nutotal}{\nu_{\text{total}}}
\newcommand{\numagic}{\nu_{\text{magic}}}
\newcommand{\nunonmagic}{\nu_{\text{non-magic}}}

\def\Weight{W}
\def\H#1{\mathrm{H}_{\text{#1}}}
\def\Vtng{V_{\text{TNG}}}
\def\Vtngind#1{(\Vtng)_{#1}}
\def\k{k_{\text{magic}}}
\def\rhoproj#1{\hat\rho_{#1}}
\def\C#1{\left(\frac{1}{C}\right)_{#1}}
\def\Auc{A_{\rm{uc}}}

\def\Vbare{V^{\text{bare}}}
\newcommand\strain{\epsilon_{\text{strain}}}

\hypersetup{colorlinks=true,linkcolor=blue,citecolor=blue,urlcolor=blue}

\newcommand{\caltechPH}{Department of Physics, California Institute of Technology, Pasadena, California 91125, USA}
\newcommand{\caltechAPH}{T. J. Watson Laboratory of Applied Physics, California Institute of Technology,
  1200 East California Boulevard, Pasadena, California 91125, USA}

\newcommand{\caltechIQIM}{Institute for Quantum Information and Matter, California Institute of Technology, Pasadena, California 91125, USA}

\begin{document}
\title{The electrostatic fate of $N$-layer moiré graphene
}

\author{Kry\vaccent{s}tof Kol\'a\vaccent{r}}
\email{Correspondence: kolar@zedat.fu-berlin.de}
\affiliation{Dahlem Center for Complex Quantum Systems and Fachbereich Physik, Freie Universit{\"a}t Berlin, 14195 Berlin, Germany}

\author{Yiran Zhang}
\affiliation{\caltechAPH}
\affiliation{\caltechIQIM}
\affiliation{\caltechPH}

\author{Stevan Nadj-Perge}
\affiliation{\caltechAPH}
\affiliation{\caltechIQIM}

\author{Felix von Oppen}
\affiliation{Dahlem Center for Complex Quantum Systems and Fachbereich Physik, Freie Universit{\"a}t Berlin, 14195 Berlin, Germany}

\author{Cyprian Lewandowski}
\address{National High Magnetic Field Laboratory, Tallahassee, Florida, 32310, USA}
\address{Department of Physics, Florida State University, Tallahassee, Florida 32306, USA}

\begin{abstract}
        Twisted $N$-layer graphene (TNG) moiré structures have recently been
        shown to exhibit robust superconductivity similar to twisted
        bilayer graphene (TBG). In particular for $N=4$ and $N=5$, the  phase diagram features a
        superconducting pocket that extends beyond the nominal full filling of
        the flat band. These observations are seemingly at odds with the canonical understanding of
        the low-energy theory of TNG, wherein the TNG Hamiltonian consists of
        one flat-band sector and accompanying dispersive bands. 
        Using a
        self-consistent Hartee-Fock treatment, we explain how the
phenomenology of TNG can be understood through an interplay of in-plane Hartree
and inhomogeneous layer potentials, which cause a reshuffling of electronic bands. We extend
our understanding beyond the case of N = 5 realized in experiment so far. We decribe how the
Hartree and layer potentials control the phase diagram for devices with N $>$ 5 and tend to preclude
exchange-driven correlated phenomena in this limit. To circumvent these electrostatic constraints, we propose a new flat-band paradigm that
could be realized in large-N devices by taking advantage of two nearly flat sectors acting together
to enhance the importance of exchange effects.       

        \end{abstract}
\maketitle

\section{Introduction}

\begin{figure}[!ht]
\begin{subfigure}{0.48\columnwidth}
\caption{}
  \includegraphics[width=1\textwidth]{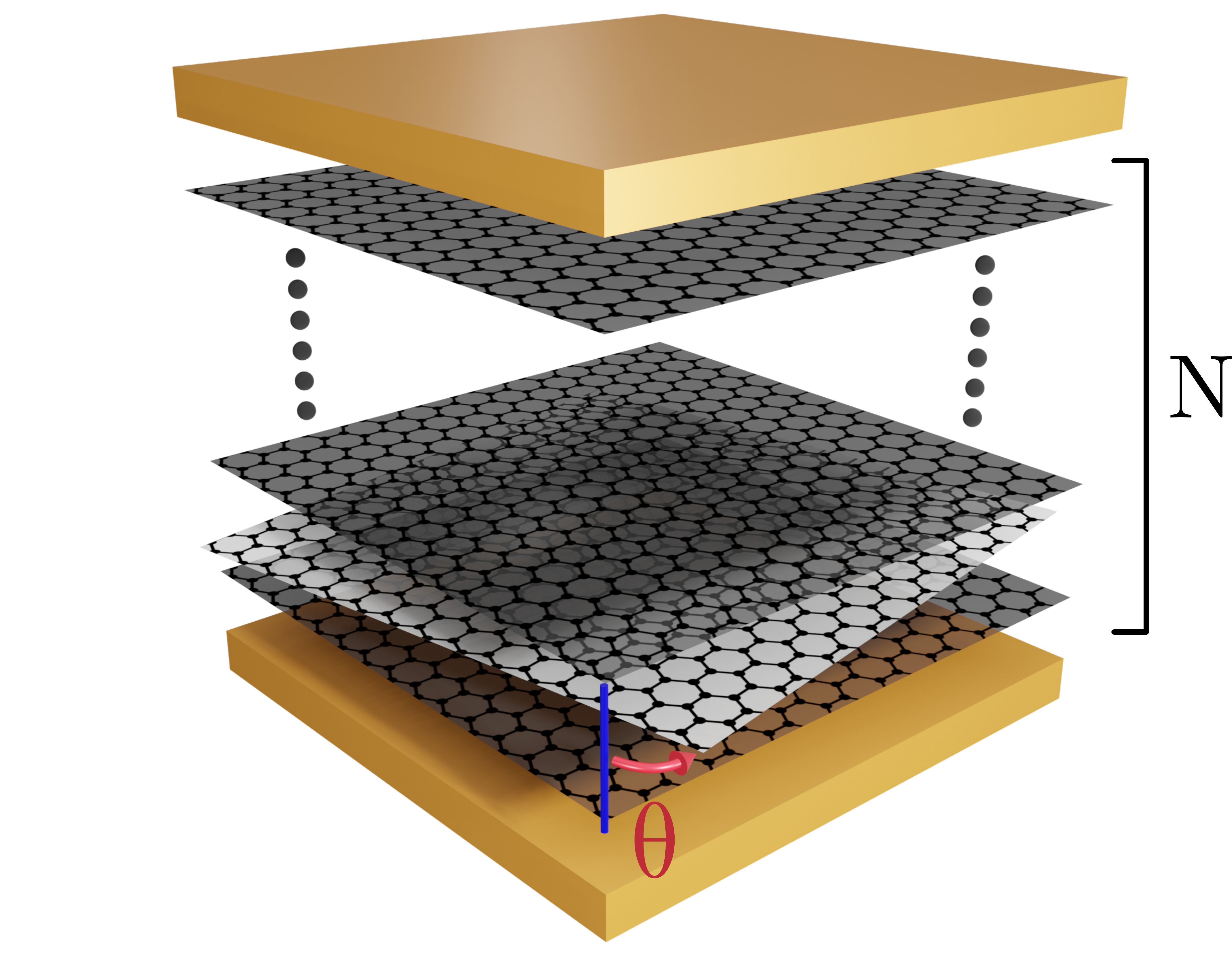}  
\end{subfigure}
\begin{subfigure}{0.48\columnwidth}
\caption{}
  \includegraphics[width=1\textwidth]{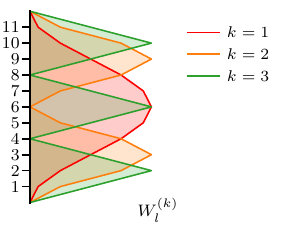}  
  \label{subfig:schematicsoflayer}
\end{subfigure}
\begin{subfigure}{1\columnwidth}
  \includegraphics[width=1\textwidth]{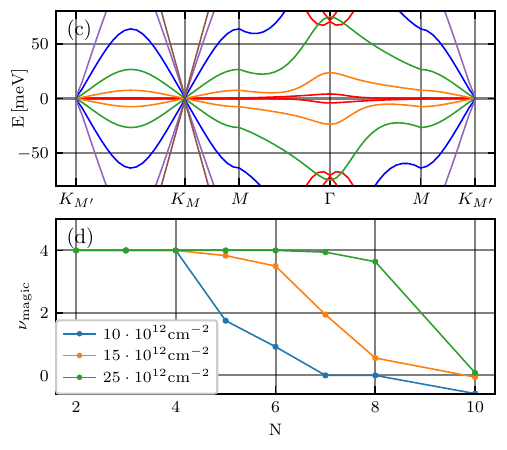}  
\end{subfigure}
\caption{
(a) Device schematics. We consider $N$-layer graphene with alternating twist angles in a double-gated setup. Here $\theta$ is the physical twist angle.
(b) Schematics of layer charge distribution (See Eq.~\eqref{eq:weightsevaluated})  for $N=11$, showing the
three sectors with lowest effective twist angle, $k=1$: red, $k=2$: orange, $k=3$: green. 
In experiment to date, $k=1$ is the flattest, ``magic" sector. 
(c) Single particle band structure for $N=11$. 
(d)  Band filling of the magic sector at different total gate densities. $25 \cdot 10^{12}\, \si{cm^{-2}}$ 
is the threshold of dielectric breakdown in current hBN-based samples
\cite{placaisPierretDielectricPermittivityConductivity2022}.}
\label{fig:fig_1}
\end{figure}

The paradigm of twisting and stacking graphene layers has led to numerous
unexpected and exciting discoveries in recent years, with the two most studied
phenomena being superconductivity and interaction-driven insulating states \cite{jarillo-herreroParkRobustSuperconductivityMagicangle2022,jarillo-herreroParkTunableStronglyCoupled2021,nadj-pergeZhangPromotionSuperconductivityMagicangle2022,deanYankowitzTuningSuperconductivityTwisted2019,kimHaoElectricFieldTunable2021,yazdaniOhEvidenceUnconventionalSuperconductivity2021,nadj-pergeKimEvidenceUnconventionalSuperconductivity2022,efetovLuSuperconductorsOrbitalMagnets2019,jarillo-herreroCaoCorrelatedInsulatorBehaviour2018,jarillo-herreroCaoUnconventionalSuperconductivityMagicangle2018}. In
part to advance a microscopic understanding of the pairing mechanism, community
efforts have focused on extending the number and types of moiré materials that
present robust superconductivity, with each device type shedding additional light on
the features 
that are conducive to pairing \cite{nadj-pergeAroraSuperconductivityMetallicTwisted2020,efetovStepanovUntyingInsulatingSuperconducting2020,youngSaitoIndependentSuperconductorsCorrelated2020,liLiuTuningElectronCorrelation2021,jarillo-herreroParkTunableStronglyCoupled2021,kimHaoElectricFieldTunable2021,jarillo-herreroCaoNematicityCompetingOrders2021}.
One extension, following the seminal prediction of
Refs.~\cite{vishwanathKhalafMagicAngleHierarchy2019,kruchkovCarrUltraheavyUltrarelativisticDirac2020}, was to stack graphene layers with
alternating twist angles, see Fig.~\ref{fig:fig_1}a. This procedure
yields a system that can be understood in terms of a set of twisted-bilayer-graphene-like bands, referred to as sectors,  at different effective
twist angles as illustrated in Fig.~\ref{fig:fig_1}c 
\cite{vishwanathKhalafMagicAngleHierarchy2019,kruchkovCarrUltraheavyUltrarelativisticDirac2020,
vishwanathLedwithTBNotTB2021}. 
The sectors (labeled by $k$) are characterized by different charge distributions over the layers, see Fig.~\ref{fig:fig_1}b. At a magic angle of the multilayer system, one sector's effective angle is just the magic angle of twisted bilayer graphene. This
prediction was realized in twisted tri-, quad-, and pentalayer moiré
materials~\cite{jarillo-herreroParkRobustSuperconductivityMagicangle2022,nadj-pergeZhangPromotionSuperconductivityMagicangle2022}. The phase diagrams observed in these experiments were
qualitatively similar to 
the phase diagram of twisted bilayer graphene. At the same time,  the experimental results for the location of the phase boundaries were  puzzling, introducing a few experimental questions. It is these questions which we address in this work.

To this end, we carry out a systematic self-consistent
Hartree-Fock analysis of interaction effects in multilayer alternating angle moiré
structures. We focus first on understanding and explaining the experimental
results of Refs.\
\cite{jarillo-herreroParkRobustSuperconductivityMagicangle2022,nadj-pergeZhangPromotionSuperconductivityMagicangle2022} 
for $N=3,4,5$ layers, and then apply the developed framework to study
the twisted $N$-layer problem (TNG, $N>5$), characterizing its electronic properties. 
Specifically, we show that as layer number
increases, larger gate voltages are required to compensate for the charge
redistribution due to interactions. This effect makes it increasingly prohibitive to electrostatically
dope $N>5$ multilayer structures into the regime where the magic flat band is
optimally filled for superconductivity as shown in Fig.~\ref{fig:fig_1}d. Interestingly, we find that while going beyond $N=5$ layers to study interaction effects of the $k=1$ flat band presents little advantage, 
focusing on the second-harmonic bands ($k=2$) in Fig.~\ref{fig:fig_1}b for $N\geq 5$ instead overcomes the prohibitive electrostatic barrier and yields
very flat bands conducive to correlation effects.

Our manuscript is structured as follows. Section II presents a summary of our results focusing on
physical understanding and experimental trends. Section III outlines the formal
description of the $N$-layer problem and introduces the Hartree-Fock machinery,
emphasizing the similarities and differences with twisted bilayer graphene (TBG).
In Sec.\ IV, we combine physical understanding with Hartree-Fock calculations for $N=3,4,5$,
focusing on explaining experimental trends. Section V discusses the electronic properties of $N > 5$ devices in more detail. We conclude with a
summary and discussion in Sec.\ VI. Readers uninterested in details
of the mathematical description can focus on Secs.\ II, V, and VI.

\section{Physical understanding and summary of main results}
\label{subsec:physicalunderstandingintro}
\subsection{Experimental motivation} 
A key physical effect in twisted alternating-angle graphene multilayers is the cascade 
of  ``resets'' close to integer fillings of the flat bands. The resets already occur at relatively high  temperatures, well  above those required for  the correlated superconducting and insulating states, and are deduced from measurements of the chemical potential
\cite{ilaniZondinerCascadePhaseTransitions2020a,yazdaniWongCascadeElectronicTransitions2020} as well as the Hall conductivity \cite{youngSaitoIndependentSuperconductorsCorrelated2020,jarillo-herreroParkTunableStronglyCoupled2021}. The cascade of transitions can be explained in different ways
\cite{ilaniZondinerCascadePhaseTransitions2020a,yazdaniWongCascadeElectronicTransitions2020,vafekKangCascadesLightHeavy2021,guineaCeaBandStructureInsulating2020,oregShavitTheoryCorrelatedInsulators2021,macdonaldXieWeakFieldHallResistivity2021,basconesDattaHeavyQuasiparticlesCascades2023}, with Ref.\ \cite{ilaniZondinerCascadePhaseTransitions2020a} interpreting it as Stoner-like 
flavor (spin and valley) polarization. Within this picture, flat-band superconductivity is
unlikely to exist when three of the four flavors are fully occupied
and time-reversed
partners at the Fermi level are absent. In
TBG, this happens beyond $\nu=\pm 3$ (see further discussion in Sec.\ III regarding intervalley coherent orders). Irrespective of the detailed theoretical
symmetry-breaking mechanism, this expectation is in line with experimental trends. In TBG, a cascade transition near  $\nu=\pm 3$
typically serves as an upper filling bound for superconductivity \cite{youngSaitoIndependentSuperconductorsCorrelated2020,nadj-pergePolskiHierarchySymmetryBreaking2022,jarillo-herreroParkTunableStronglyCoupled2021,kimHaoElectricFieldTunable2021}. Similarly, a lower filling bound for superconductivity is the cascade transition at $\nu=\pm 2$.

Cascade phenomenology has also been reported for TNG systems with $N=3,4,5$ layers \cite{jarillo-herreroParkTunableStronglyCoupled2021,kimHaoElectricFieldTunable2021,jarillo-herreroParkRobustSuperconductivityMagicangle2022,nadj-pergeZhangPromotionSuperconductivityMagicangle2022}, c.f., Fig.~\ref{fig:fig_2}a. The band structure of TNG decomposes into decoupled sectors of TBG-like and (for $N$ odd) monolayer-graphene (MLG)-like bands. This is illustrated by the  example band structure for $N=11$ in Fig.~\ref{fig:fig_1}c, which features five TBG-like bands and a Dirac cone. When one of the TBG-like sectors is  effectively at the magic angle, the cascade features can  be understood as occurring in the magic sector, with the other sectors being filled uniformly~\cite{jarillo-herreroParkRobustSuperconductivityMagicangle2022,nadj-pergeZhangPromotionSuperconductivityMagicangle2022}.

Startlingly, as shown in Fig.~\ref{fig:fig_2}b and reported in Refs.~\cite{jarillo-herreroParkRobustSuperconductivityMagicangle2022,nadj-pergeZhangPromotionSuperconductivityMagicangle2022}, superconductivity persists to higher total fillings in TQG ($N=4$) and TPG ($N=5$),  extending up to $\nu=5$ for the $N=5$ case of TPG. Simultaneously the cascade ``resets'' also set in at higher filling fractions. Assuming that the magic sector is in the optimal doping range for superconductivity, i.e., it has 2-3 electrons per moiré cell, these observations would imply substantial filling of the nonmagic sectors at odds with a simple band-structure picture. The nonmagic sectors are strongly dispersive,
so that their noninteracting band structure would
predict almost no filling. Specifically, complete filling of the magic bands would be 
accompanied by less than ${\sim} 0.06$ electrons per unit cell in the nonmagic bands for TPG and less than ${\sim} 0.02$ for TQG. 

To obtain these estimates, we note that 
the $\lfloor N/2 \rfloor$ TBG-like electronic sectors ($k=1,2,\dots,\lfloor N/2 \rfloor$) have effective twist angles  \cite{vishwanathKhalafMagicAngleHierarchy2019}
    \begin{equation}
    \label{eq:heuristicthetaeffdef}
    \thetaeff{k}= 
    \frac{\theta}{2 \cos[\frac{\pi k}{N+1}]}\,,
    \end{equation}
which differ from the physical twist angle $\theta$, Fig.~\ref{fig:fig_1}a. 
This formula reveals the advantage of multilayers -- one 
can obtain a sector effectively at the magic angle while actually lying at a larger twist angle.
This is best exploited by choosing the $k=1$ sector to lie effectively at the magic angle,
which maximizes the physical twist angle. All the current experiments on TNG make this choice, and
we shall also make it our default choice for analysis.
However, we note that for large $N$, the choice $\k=2$ also becomes feasible. We will return to this possibility in Sec.\ \ref{sec:largenanalysis}.
Approximating the nonmagic sectors as Dirac cones, their filling is (see App.~\ref{subsections:dos_non_int_dirac_cone})
\begin{equation}
\label{eq:formulanuk}
\nunonmagic =\sum_{k\in \text{nonmagic}} \nu_k \approx \sum_{k\in \text{nonmagic}} 
\frac{A_{\text{uc}}N_f c_k}{4\pi (\hbar v_D^{(k)})^2} \mu_k^2.
\end{equation}
Here, $c_k=2$ ($c_k=1$) if the sector $k$ is TBG-like (MLG-like),
$\mu_k$ is the effective chemical potential in 
sector $k$, $N_f =4$ is the number of flavors,
and $v^{(k)}_D$ is the Dirac velocity in sector $k$. In the absence of interactions, $\mu_k=\mu_{\text{magic}}$ with $\mu_{\text{magic}}$ the magic sector Fermi energy. A filled magic sector corresponds to $\mu_{\text{magic}} \approx W/2$, where $W$ is the noninteracting bandwidth. 
This bandwidth varies with strain, taking values  $W \approx 2-20 \, \si{meV}$.
Even at the upper limit for $W$, we then find $\nunonmagic \lesssim 0.06$ for TPG (using $v_D^{(k=2)} \approx 0.35 v_D$). For TQG, the $k=2$ sector has an even larger detuning from the magic angle 
($\thetaeff{k=2}=2.9^\circ$), so that  $v_D^{(k=2)} \approx 0.6v_D$ and $\nunonmagic \lesssim 0.02$. Therefore, the enhanced nonmagic-sector filling \cite{jarillo-herreroParkRobustSuperconductivityMagicangle2022,nadj-pergeZhangPromotionSuperconductivityMagicangle2022} is an interaction effect, motivating 
our Hartree-Fock study of TNG. 

\subsection{Physical understanding}

Electron-electron interactions alter the above considerations predominantly through two terms in the Hamiltonian, as can be seen by examining the mean-field decomposition
\begin{equation}
\label{eq:hmfcomplete}
\H{MF}=\H{SP} + \H{Hartree}+ \H{Fock}+ \H{layer}.
\end{equation}
First, interactions represented by the Hartree and Fock mean-field terms broaden the noninteracting magic bands, promote the onset of symmetry-breaking order, and, crucially for our analysis, induce  filling-dependent upward shifts of the quasiparticle energies relative to nonmagic sectors. This Hartree-dominated shift 
arises because the electron density of the TBG-like sectors is spatially inhomogeneous in the 2D plane, which is associated with a cost in Coulomb energy
\cite{waletGuineaElectrostaticEffectsBand2018,guineaCeaElectronicBandStructure2019,guineaCeaBandStructureInsulating2020,melladoRademakerChargeSmootheningBand2019,lischnerGoodwinHartreeTheoryCalculations2020}.
Importantly, the inhomogeneity is particularly strong in the magic sector and decreases with detuning from the magic angle. Second, the contribution $\H{layer}$ is new to $N>2$ layers and arises because the sectors have different vertical charge distributions across layers \cite{jarillo-herreroParkRobustSuperconductivityMagicangle2022,nadj-pergeZhangPromotionSuperconductivityMagicangle2022}
 as shown in Fig.~\ref{fig:fig_1}b.
These distributions are given by the layer dependence of the wave functions, taking the form of standing waves analogous to a particle-in-a-box problem. 
The sector with lowest effective twist angle, $k=1$, corresponds to the first harmonic, which is singly peaked at the center of the stack. The $k=2$ sector is the second harmonic with a doubly-peaked structure, and so on. The different layer-dependent charge distributions imply that the sectors have 
different energies due to the electric potential produced by the gate charges.

For the devices investigated experimentally (magic sector $k=1$), both $H_{\mathrm{Hartree}}$ and $H_{\mathrm{layer}}$ effects
enhance the occupation of the nonmagic sectors relative to the noninteracting
band-structure scenario described above. The first mechanism postpones the occupation
of the magic sector as it is broadened and shifted upward in energy as it is
filled. A similar shift in energy also occurs for the second mechanism. The
potential produced by the gate charges in combination with the induced charges
in TNG has a maximum in the central layer. (Note that in the absence of a
displacement field, the electric field vanishes at the center by symmetry.
Moreover, the potential drops towards the, say, positively charged gate electrodes above and below the TNG stack.) Due to this potential maximum, the
energy is higher for sectors, in which charge is more localized near the
central layer. Thus, this mechanism also predicts that the magic sector is
pushed up in energy relative to the nonmagic sectors.

In subsequent sections, we quantify these effects by extensive Hartree-Fock calculations, but for the purpose of developing physical understanding we can nonetheless arrive at some analytical results that qualitatively reproduce numerical trends. To do so, we make the rough approximation that as a result of interactions the overall bandstructure of each sector remains fixed (i.e. given by the noninteracting band structure) and only the chemical potential of each sector $\mu_{k}$ shifts as
\begin{equation}
    \mu_{k} = \mu-U_k-G_k\,.
\end{equation}
Here, $U_k$ and $G_k$ quantify the shifts due to $H_\text{layer}$ and $\H{Hartree}$, and $\mu$ is the chemical potential of the whole system.
We take $G_k=0$ for all sectors except the magic sector ($k=1$) as it has the largest in-plane inhomogeneity (see App.~\ref{app:spHartreetwistnagledep}).
In the magic
sector \cite{waletGuineaElectrostaticEffectsBand2018,guineaCeaElectronicBandStructure2019,guineaCeaBandStructureInsulating2020},
$G_{k} \sim e^2/(4\pi \epsilon_{\parallel}\epsilon_0 L_M)$, where $L_M$ is the
moiré period. For TPG, depending on dielectric constant, 
we estimate $\SI{10}{meV} \lesssim G_{k} \lesssim \SI{30}{meV}$. For full filling of the
magic sector, $\mu_{1}{\sim} W/2{\sim}2 - 10$ meV, an extension of the noninteracting analysis above gives $\mu_{2} \approx 12-\SI{40}{meV}$, yielding $\nu_{\text{non-magic}}\approx 0.1 - 1.1$.
[Note that for $\nu_{\text{non-magic}} \gtrsim 0.5$, we need to employ the full density of states, which deviates from the Dirac approximation in Eq.~\eqref{eq:formulanuk}].

Inclusion of the shift $U_k$ induced by $H_\text{layer}$
can further increase the filling of $\nu_\text{non-magic}$. At the mean-field level, the charge distribution 
across the layers enters the Hamiltonian through 
 \begin{equation}
 \label{eq:hmfperp}
\H{layer} = -e \sum_l {\rhoproj{l,0}} V_l,
\end{equation}
where $V_l$ is the potential and  $\rhoproj{l,0}$ the electron number (i.e., the $\mathbf{q}=0$ Fourier component of the electron density $\hat\rho_{l,\v{q}}$) of layer $l$. The term $\H{layer}$ contributes nontrivially due to imperfect screening of the gate electrodes by the layers 
and becomes increasingly important as $N$ grows. 
In the absence of interaction-induced sector mixing, we consider the energy shift 
\begin{equation}
\label{eq:capacitanceformula}
 U_{k} = e^2 \frac{d_l}{\Auc\epsilon_0\epsilon_\perp} 
\sum_{k'} (C^{-1})_{k,k'} \nu_{k'}
\end{equation}
of a sector $k$ for given sector fillings $\nu_k$. Here, $A_{\rm{uc}}$ is the unit-cell area, $d_l$ is the layer distance, and $\epsilon_\perp$ the out-of-plane dielectric constant of the graphene layers.
The matrix $C$ in sector space is a capacitance-like matrix, made dimensionless by extracting an appropriate prefactor. 
This matrix succinctly accounts for the charge distributions of the sectors
over the layers. We tabulate $C^{-1}$ for TQG, TPG, as well as large $N$ in
Table~\ref{tab:ckkprime} (see App.~\ref{app:analyticlayerpotentials} for formulas for arbitrary $N$ and derivations).

\begin{table}[t]
\begin{tabular}{c|c|c|c}
$N$ & $(C^{-1})_{1,1}$ & $(C^{-1})_{1,2}$ = $(C^{-1})_{2,1}$ &$(C^{-1})_{2,2}$ \\
\hline
4  & $0.262$  & $0.1$ & $0.0382$\\
\hline
5  & $0.403$ & $0.208$   & $0.125$  \\ 
\hline
$N \to \infty$  & $0.147\, N $  & $0.115\, N $ &$0.099\, N $
\end{tabular}
\caption{Inverse capacitance $(C^{-1})_{k,k'}$ for $k,k' \in \{1,2\}$, evaluated for layer numbers $N=4$, $N=5$, and $N\to \infty$. }
\label{tab:ckkprime}.
\end{table}

With the help of Table \ref{tab:ckkprime}, we obtain that the effective chemical potential in
the nonmagic TBG-like sector increases by
\begin{equation}
\label{eq:analytical_potential}
U_1-U_2
= 
\frac{e^2d_l}{A_{\rm{uc}}\epsilon_0\epsilon_\perp}
\left \{  
\frac{7}{36}\,\, \numagic +
\frac{3}{36}\,\, \nu_{k=2}  
\right\}\,,
\end{equation}
where the numerical coefficients are $(C^{-1})_{1,1}-(C^{-1})_{2,1}$ and
$(C^{-1})_{1,2}-(C^{-1})_{2,2}$, respectively. 
This shows explicitly that layer
potentials increase the effective chemical potential of the nonmagic sector
and hence the filling, as can be obtained from Eq.~\eqref{eq:formulanuk}. Plugging in numbers for $\numagic = 4$,  
we obtain a $7 -\SI{45}{meV}$ shift (for $\epsilon_\perp \in [2,12]$;
see Sec.~\ref{sec:coulombint} for a discussion 
of the role of dielectric constants). Including only the single-particle term $\H{SP}$ and $\H{layer}$, the effective chemical potential in the TBG-like sector is $\mu_2=W/2 + U_1-U_2 \approx 9-\SI{55}{meV}$, 
corresponding to $\nunonmagic \approx 0.05 - 2.5$. 
At small nonmagic fillings, the effect of the layer potential $\H{layer}$ is reinforced by the Hartree correction $\H{Hartree}$.
In fact, due to the linear density of states of the nonmagic Dirac cones, the nonmagic filling depends nonlinearly on $\H{layer}$ and $\H{Hartree}$, cf.~Eq.~\eqref{eq:formulanuk}.

To illustrate this, consider $U_1-U_2 = \SI{10}{meV}$ and $G_1 = \SI{10}{meV}$ and small bandwidth $W/2=\SI{2}{meV}$. 
Taken separately, each term would only yield a tiny $\nunonmagic \sim 0.07$.
On the other hand, taking $\mu_2 = \SI{22}{meV}$ in
Eq.\ \eqref{eq:formulanuk} yields a four times larger $\nunonmagic  \sim 0.3$.
This highlights the importance of considering both shift mechanisms.

The inverse capacitance matrix $(C^{-1})_{k,k'}$ is a decreasing function of $k$ and $k'$. Physically, larger-$k$ sectors screen the gate field better, therefore generating smaller potentials.
This monotonic decrease implies that 
$U_\text{magic} - U_k>0$ for any (nonmagic) $k>1$. Thus, the $\mu_k$ of nonmagic sectors increases, 
enhancing their occupations.
Secondly, for fixed $k$ and $k'$, $(C^{-1})_{k,k'}$ scales linearly with the vertical extent (as the inverse capacitance of a parallel-plate capacitor) and thus with the number of layers $N$. This suggests that the layer potential grows in importance with $N$, eventually dominating over other contributions for large $N$. 
Indeed, other contributions to the mean-field Hamiltonian do not scale with the number of layers.
This suggests that the layer potentials become dominant at large $N$ and doping of the central $k=1$ 
sector by gating will be preempted by dielectric breakdown \cite{placaisPierretDielectricPermittivityConductivity2022}, as shown in Fig.~\ref{fig:fig_1}d. We return to this analysis using Hartree-Fock calculations in subsequent sections. 

\section{Model}
\label{sec:model}
In this section, we introduce the  noninteracting model, specify the 
interaction, and discuss the mean-field decoupling. 
While we largely follow standard procedures for the mean-field description of moir\'e graphene  \cite{guineaCeaBandStructureInsulating2020,vishwanathLiuNematicTopologicalSemimetal2021,zaletelBultinckGroundStateHidden2020,lianXieTwistedSymmetricTrilayer2021,scheurerChristosCorrelatedInsulatorsSemimetals2022,bultinckKwanKekulSpiralOrder2021,parameswaranWagnerGlobalPhaseDiagram2022}, 
we allow for layer dependence of the interaction and include the layer potential term that is usually ignored.

\subsection{Twisted graphene multilayers}
We consider  $N$-layer alternating angle twisted graphene. Focusing on the K-valley, the single-particle Bistritzer-MacDonald Hamiltonian reads \cite{vishwanathKhalafMagicAngleHierarchy2019}
\begin{widetext}  
\begin{equation}
\label{eq:spham}
H^{K}_{\text{sp}}= \begin{pmatrix}h_{-\theta/2}(\v k) & T^\dagger(\v r) & 0&\cdots& 0 \\ 
T(\v r)  &h_{\theta/2}(\v k) & T(\v r) & &  \\
0 & T^\dagger(\v r)  &h_{-\theta/2}(\v k) &    &\\
\vdots &   & &  \ddots   & \\
0&  & &     & h_{(-1)^N \theta/2}(\v k) \\
\end{pmatrix},
\end{equation}
\end{widetext}
where $h_{\theta/2}(\v k) = - i \hbar v_D (\boldsymbol{\sigma} \cdot \v k)
e^{i\theta\sigma_z}$ denotes the Dirac Hamiltonians of the layers (our numerics neglects the rotation of the Dirac terms)
and $T(\v r) = \sum_{j=0}^2 T_j e^{i\v q_j\cdot \v r}$ is the interlayer hopping with $T_j = w_{AA}\sigma_0 + w_{AB} \left[\sigma_x \cos(2\pi j/3)+
\sigma_y \sin(2\pi j/3)\right] $ and $\v q_j = (O_3)^j (\mathbf{K}_2-\mathbf{K}_1) = 2|K| \sin\left(\theta/2\right) (O_3)^j  \left[0 ,-1\right] $ with $\mathbf{K}_i$ the Dirac-point positions in layer $i$ 
and $O_3$ the matrix of a counterclockwise $120^\circ$ rotation.
Neglecting possible layer dependence, 
we account
for lattice relaxation by choosing $w_{AA}= 80 \si{meV} , w_{AB} = 110 \si{meV}$ \cite{vishwanathLedwithTBNotTB2021}.
Dispersion and Bloch wave functions of the
$K'$-valley follow by time-reversal
symmetry. The model of Eq.~\eqref{eq:spham} is a minimal description of $N$-layer systems, neglecting relative layer displacements \cite{vishwanathKhalafMagicAngleHierarchy2019,macdonaldQinInPlaneCriticalMagnetic2021}, 
next-nearest-layer hoppings
\cite{vishwanathKhalafMagicAngleHierarchy2019},
periodic strain \cite{namKoshinoEffectiveContinuumModel2020}, and layer
dependence of lattice corrugation \cite{vishwanathLedwithTBNotTB2021}. 
While these additional ingredients modify the quantitative details of the electronic
spectrum, they do not alter the two key features, namely the inhomogeneous charge
distribution and the
inhomogeneous distribution of electronic sectors across layers. Both ingredients are crucial to capture the effect of interactions on the properties of the $N$-layered structure.

The single-particle Hamiltonian $H_\mathrm{sp}$ transforms into 
block-diagonal form 
under a basis transformation $\Vtng$ in layer space \cite{vishwanathKhalafMagicAngleHierarchy2019}. For an even number $N$ of layers, there are $N/2=\lfloor N/2 \rfloor$ blocks -- or sectors. These blocks describe bands analogous to twisted bilayer graphene at twist angle $\theta$ with interlayer hoppings rescaled by a coefficient $\Lambda_k$. 
We can equivalently think of the sectors as corresponding to TBG with unscaled
hoppings, but an effective twist angle 
\begin{equation}
\label{eq:thetaeffdef}
\thetaeff{k} = \theta / \Lambda_k.
\end{equation}
In this picture, the sector Hamiltonian is multiplied by an overall scale factor $\Lambda_k$. For $N$ odd, in addition to 
the $\lfloor N/2\rfloor$ TBG-like sectors,
there is an additional sector, in which the band derives from the underlying
graphene Dirac cone folded into the moir\'e Brillouin zone (BZ). We will
denote this sector as the monolayer-graphene (MLG)-like sector (see Fig. \ref{fig:fig_1}b). 
We will choose 
the physical angle $\theta$ such that
there is one TBG-like sector -- termed magic sector -- at the magic angle, $\thetaeff{k} = \theta^{\text{magic}} \approx 1.1^\circ$. 
In experiments to date, this would be the $k=1$ sector, but in Sec.\ \ref{sec:largenanalysis} we also consider the possibility $\k=2$.
We refer to all other sectors as the nonmagic sectors, including the MLG-like sector for $N$ odd \cite{vishwanathKhalafMagicAngleHierarchy2019}. 
Technically, the sector decomposition emerges by solving an effective $\lfloor
N/2 \rfloor$-site open tight-binding chain on the even layers, with $\lfloor
N/2 \rfloor$ solutions (see App.~\ref{appsubsec:spphysicssectordecomposition} for a pedagogical derivation). 
The solution for the odd layers proceeds analogously. The resulting weight
distribution for sector $k$ is \begin{equation} \label{eq:weightsevaluated}
\Weight^{(k)}_l   = \frac{2}{N+1} \sin^2\left( \frac{\pi kl}
{N+1}\right),\end{equation} as plotted in Fig.~\ref{fig:fig_1}c, with
corresponding  eigenvalues \begin{equation}    \label{eq:lambdaksol}\Lambda_k =
2 \cos\left(\frac{  \pi k}{N+1}\right).\end{equation} Combined with Eq.\
\eqref{eq:thetaeffdef}, this gives Eq.\ \eqref{eq:heuristicthetaeffdef}.  With
an increasing number of layers, there is a continuum of twist angles
\cite{vishwanathKhalafMagicAngleHierarchy2019} , with the largest density of
twist angles  close to the minimal $\thetaeff{k}$ (attained for $k=1$). If the
physical twist angle $\theta$ is such that the lowest effective angle sector is
magic, there will thus be other sectors very close to the magic angle.
Moreover, by slightly decreasing the physical twist angle, one can
alternatively tune the larger effective angles to be magic (see Sec.~\ref{sec:largenanalysis}). The weight distribution in Eq.~\eqref{eq:weightsevaluated} quantifies the charge distributions across layers
for the various sectors, see Fig.~\ref{fig:fig_1}c. As discussed above, this is
important for the electrostatic properties of the problem.

\subsection{Coulomb interactions}
\label{sec:coulombint}

We assume a symmetric double-gated setup (see Fig.\ \ref{fig:fig_1}a) as typically employed in experiment. We work at gate charge densities $e n/2$ per gate, so that $-en$ is the charge density in TNG. We include
Coulomb interactions through
\begin{equation}
\H{int} = \frac{1}{2}\int d\v r d\v r' V(\v r-\v r') :\rho(\v r)\rho (\v r'):,
\end{equation}
where the density $\rho(\v r)$ includes free charges in both the graphene system and on the gates with the positive background subtracted. The integration ranges over the full 3D space. Integrating out the electronic degrees of freedom of the metallic gates, one arrives at an effective screened interaction for the $N$ layers for a fixed
electron density $n$ (see App.~\ref{app:vbaretoveff} for details). The resulting interaction Hamiltonian takes the form
\begin{eqnarray}
\H{int} &=& \frac{1}{2A} \sum_{\v{q}\neq 0}\sum_{i,j} V_{ij}(\v q) :\rho_{i,\v q} \rho_{j,-\v q}:
\nonumber\\
&&\qquad \qquad +\sum_{i=1}^{N-1} A \epsilon_\perp \epsilon_0 d_l\frac{(E^{\perp \,}_{i,i+1})^2}{2}.
\label{eq:interacting_hamiltonian}
\end{eqnarray}
Here, $\rho_{i,\v q}$ is the electron density in layer $i$ at in-plane momentum $\v q$, $E^\perp_{i,i+1}$ denotes the uniform component of the perpendicular electric field between layers $i$ and $i+1$, $A$ is the system area, $d_l$ is the interlayer distance, and
$V_{ij}(\v q)$ is the double-gate-screened layer-dependent Coulomb interaction derived in App.~\ref{app:layerdepcoulombV}. 
We allow the dielectric constant of the $\v q = 0 $ term 
($\epsilon_\perp$) to differ from the dielectric constant 
entering $V_{ij}(\v q)$ ($\epsilon_\parallel$). 
Physically, the out-of-plane interaction reflects the out-of-plane response of graphene,
while the $\v q \neq 0$ component is governed by the dielectric properties of the substrate.
For graphene layers, $\epsilon_\perp$ has been estimated to be around $2$  
\cite{guineaGuineaChargeDistributionScreening2007,serbynGhazaryanMultilayerGraphenesPlatform2022}, 
while $\epsilon_\parallel$ is around $5$ for hBN substrates
\cite{vandenbergheLaturiaDielectricPropertiesHexagonal2018,guineaCeaElectronicBandStructure2019,lianBernevigTBGIIIInteracting2021,vafekKangStrongCouplingPhases2019}.
Larger values, accounting for remote band screening, have also been investigated 
\cite{macdonaldXieNatureCorrelatedInsulator2020, vishwanathLiuNematicTopologicalSemimetal2021,guineaCeaElectronicBandStructure2019}.
We treat the dielectric constants as parameters.
Without the second term, Eq.\ \eqref{eq:interacting_hamiltonian} is the standard in-plane
Coulomb interaction of a 2D system with screening due to metallic gates. The second term is not usually included, but is important for multilayer systems as discussed in Sec.\ \ref{subsec:physicalunderstandingintro}. 

\subsection{Mean-field decoupling}
We perform our numerical calculations 
by restricting
the full Hilbert space to a 
finite number of $N_{\text{active}}$ bands with
$N_{\text{flavor}}$ spin/valley flavors
and solving the mean-field Hartree-Fock
equations. We  search for the 
 $N_{\text{active}}\times N_{\text{active}}$ density matrix 
$[P_f(\v k)]_{\alpha \beta} = \langle 
c^\dagger_{f,\v k,\alpha }c^{\phantom{\dagger}}_{f,\v k,\beta}  
\rangle$. Here $c_{f,\v k,\beta }$ annihilates a flavor-$f$ electron
in the single-particle band $\beta$ at momentum $\v k$. The single-particle bands fall into sectors
$k \in \{1, \ldots, n_o\}$. We keep $N_{\text{remote}}$ remote bands, which generate additional Hartree and Fock interaction terms. In projecting onto a finite set of bands, we are assuming
frozen fully filled bands below and empty bands above this set. To avoid overcounting of interactions already present in monolayer graphene and thus included in the BM model
\cite{macdonaldXieNatureCorrelatedInsulator2020,zaletelBultinckGroundStateHidden2020,parameswaranWagnerGlobalPhaseDiagram2022}, 
we subtract a mean-field Hamiltonian corresponding to a  
reference density matrix $P_f^0(\v k)$.
This is implemented in the mean-field
equations by replacing every  $P_f(\v k)$ with
\begin{equation}
\delta P_f(\v k) = P_f(\v k) - P^0_f(\v k).
\end{equation}
We choose the subtraction scheme \cite{macdonaldXieNatureCorrelatedInsulator2020,zaletelBultinckGroundStateHidden2020,bultinckKwanKekulSpiralOrder2021,parameswaranWagnerGlobalPhaseDiagram2022} 
in which $P_f^0(\v k)$ is the ground density matrix at
charge neutrality with the interlayer hoppings switched off.
For bands far below the charge-neutrality point, interlayer hoppings are ineffective and this
density matrix  approximates that of 
fully filled TNG bands. It therefore cancels with the remote-band-interaction 
term to a good approximation \cite{zaletelBultinckGroundStateHidden2020,lianBernevigTBGIIIInteracting2021},
justifying retaining only a finite number  $N_{\text{remote}}$ of remote bands.

For the in-plane term, the mean-field decoupling extends the usual 
procedure detailed in previous 
 studies 
\cite{guineaCeaBandStructureInsulating2020,vishwanathLiuNematicTopologicalSemimetal2021,zaletelBultinckGroundStateHidden2020,lianXieTwistedSymmetricTrilayer2021,scheurerChristosCorrelatedInsulatorsSemimetals2022,macdonaldXieNatureCorrelatedInsulator2020,vishwanathKhalafMagicAngleHierarchy2019} 
to include the layer dependence of $V_{i,j}(\v{q})$. The resulting Hartree term reads \begin{equation}
 \label{eq:hmfhartree}
\H{Hartree} =\frac{1}{A}\sum_{i,j}\sum_{\v G}\rhoproj{i,\v G} V_{i,j}(\v G) \left\langle 
\rhoproj{j,-\v G}\right \rangle,\end{equation}
where we introduce the projected layer density operator,
$\rhoproj{i,\v G} = \sum_{f \v k} c^\dagger_{f,\v k} 
\Lambda^{fi}_{\v G}(\v k) c_{f,\v k}$,
and denote the mean-field density operator 
(with the appropriate subtraction) 
as 
\begin{eqnarray} 
\langle \rhoproj{j,- \v G}\rangle  &=&\sum_{f}\sum_{\v k}
\langle c^\dagger_{f,\v k} \Lambda^{fj}_{-\v G}(\v k) c_{f,\v k}\rangle 
\nonumber\\
&=&
\sum_{f}\sum_{ \v k}\tr \left[\delta P^T_f(\v k) \Lambda^{fj}_{-\v G}(\v k)\right].
\end{eqnarray}
Here, 
the trace runs over the space of active bands. Similarly, the Fock term reads 
\begin{align}
 \label{eq:hmffock}
\H{Fock} &=-\frac{1}{A}\sum_{f}\sum_{i,j}\sum_{\v q,\v k}  
V_{ij}(\v q) 
\nonumber \\ 
&\times c^\dagger_{f,\v k}\left[
\Lambda^{fi}_{\v q}(\v k)\delta P^T_f(\v k + \v q) \Lambda^{fj}_{- \v q}(\v k +\v q)
\right]c_{f,\v k},
\end{align}
where in contrast to the Hartree term, each flavor interacts only with itself. 

App.~\ref{app:hartreefockderivation} details a formal derivation of $\H{layer}$ in Eq.~\eqref{eq:hmfperp} by decoupling the out-of-plane term in Eq.~\eqref{eq:interacting_hamiltonian}.
The difference of layer potentials
\begin{equation}
V_{i+1} - V_{i} =  - d_l E^\perp_{i,i+1}
\end{equation}
is related to the electric field, which is given by (Gauss law)
\begin{equation}
\label{eq:gausslaw}
E^\perp_{i,i+1} = -\frac{e}{\epsilon_0 \epsilon_\perp}\left\{
\frac{1}{A} \sum_{l=1}^i \langle \rhoproj{l,0}\rangle-\frac{n}{2} \right\}.
\end{equation}
We fix the arbitrary constant of $V_{i}$ by setting $V_1 + V_N = 0$.

We note in passing that Ref.~\cite{vishwanathLedwithTBNotTB2021} similarly considers interaction effects on the electronic spectrum of $N>3$ systems. The nonmagic sectors are described as a set of equal Dirac cones with the chemical potential set by that of the flat bands. Their role in the mean-field calculation is reduced to providing static RPA screening for the magic sector as given by Refs.\ \cite{guineaWunschDynamicalPolarizationGraphene2006,dassarmaHwangDielectricFunctionScreening2007}. This procedure focuses solely on describing interaction effects in the magic bands, but misses the impact of the nonmagic sectors on hybridizing the sectors and shifting their relative energies with the concomitant changes in filling. 

Our analysis assumes that the symmetry
breaking preserves the flavor index, precluding intervalley coherent states
\cite{vafekKangStrongCouplingPhases2019,zaletelBultinckGroundStateHidden2020,bultinckKwanKekulSpiralOrder2021,bultinckParkerStraininducedQuantumPhase2021}
, which are likely
the actual ground states \cite{yazdaniNuckollsQuantumTexturesManybody2023,feldmanYuSpinSkyrmionGaps2022} of twisted bilayer graphene 
\cite{macdonaldPotaszExactDiagonalizationMagicAngle2021,regnaultXieTBGVIExact2021,zaletelWangKekulSpiralOrder2022}. This limits  our analysis to qualitative features of the phase diagram of $N$-layer alternating twisted bilayer graphene. This approach has been shown to
reproduce experimental trends
\cite{ilaniZondinerCascadePhaseTransitions2020a,oregShavitTheoryCorrelatedInsulators2021}. As we will see, the phase diagram of TNG is mainly controlled by the interplay of the in-plane Hartree and layer potentials, which on the moiré scale, are insensitive to the subtle details of
flavor-symmetry breaking
\cite{guineaCeaBandStructureInsulating2020}. We thus expect our results to apply even when different candidate
ground states \cite{vafekKangStrongCouplingPhases2019,zaletelBultinckGroundStateHidden2020,bultinckKwanKekulSpiralOrder2021,bultinckParkerStraininducedQuantumPhase2021}
(such as intervalley coherent states) are considered for the magic sector. 

Experimental samples are, to some extent, always strained
\cite{bediakoKazmierczakStrainFieldsTwisted2021,yazdaniXieSpectroscopicSignaturesManybody2019,andreiJiangChargeOrderBroken2019,pasupathyKerelskyMaximizedElectronInteractions2019,nadj-pergeChoiElectronicCorrelationsTwisted2019,nadj-pergeKimEvidenceUnconventionalSuperconductivity2022}. Strain increases the kinetic energy of the bands, suppressing
interaction effects, and breaks $C_3$ symmetry, preventing gap opening
by $C_2 T$ symmetry breaking. We incorporate strain as a constant vector potential, which alternates between layers (heterostrain \cite{fuBiDesigningFlatBands2019}) as described in App.~\ref{app:strain}. This simplified  description of strain is sufficient to
capture the broadening
of the noninteracting bands as well as the $C_3$ symmetry breaking. 
Not considering intervalley coherence, we also preclude the incommensurate-Kekul\'e-spiral state \cite{bultinckKwanKekulSpiralOrder2021,parameswaranWagnerGlobalPhaseDiagram2022}, for which there is some experimental support
\cite{yazdaniNuckollsQuantumTexturesManybody2023,nadj-pergeKimImagingIntervalleyCoherent2023}.
Again, this is justified since electrostatic effects have larger
energy scales and contribute over a wider temperature range.

 \begin{figure*}[t]
  \includegraphics[width=1\textwidth]{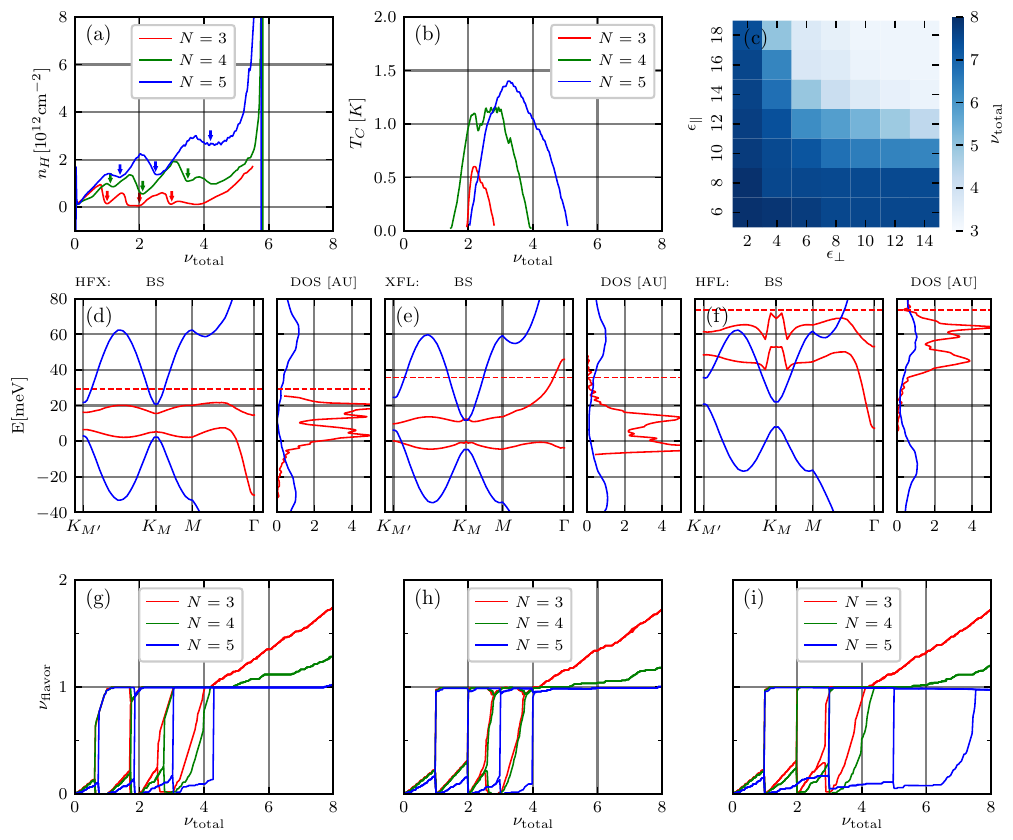}  
\caption{
(a) Experimental data of the Hall density vs.\ total filling showing the the cascade transitions (arrows) for $N=3,4,5$, see   Ref.~\cite{nadj-pergeZhangPromotionSuperconductivityMagicangle2022} for details on the samples and measurements. 
(b) Corresponding experimental data for $T_C$ domes for $N=3,4,5$. 
(c) Colormap of $\nutotal$ needed to reach filling $\numagic=3$ of the magic sector 
in the $\epsilon_\parallel$-$\epsilon_\perp$ plane.
(d) Interacting structures and densities of states for TPG at $\nutotal=4$, including in-plane Hartree and Fock (HFX) terms from Eq. \ref{eq:hmfcomplete}. 
Shown are the $k=1$ magic sector (red) and $k=2$ nonmagic TBG-like sector (blue).
(e) Same as (d)  but with layer Hartree potentials and Fock (XFL)
(f) Same as (d) but including all terms, that is, HFL.
(g) Flavor-resolved magic sector filling showing the cascade with in-plane Hartree and Fock (HFX) for $N=2,4,5$.
(h) Same as (g), but with out-plane Hartree and Fock (XFL)
(i) Same as (g), but including all the terms (HFL).}
\label{fig:fig_2}
\end{figure*} 

\section{Mean-field results for $N\leq 5$}
\label{sec:cascadeofphasetransitionsttgtqgtpg}

We now apply the mean-field approach detailed above to alternating twisted $N$-layer structures with $N=3,4,5$, confirming the qualitative reasoning discussed in Sec.~\ref{subsec:physicalunderstandingintro}. 
Figures~\ref{fig:fig_2}a,b show experimental results for the  filling dependence of the Hall density and of the superconducting $T_C$, respectively. Taken together, these data
indicate a substantial filling of the nonmagic sectors. As originally proposed in
Refs.~\cite{jarillo-herreroParkRobustSuperconductivityMagicangle2022,nadj-pergeZhangPromotionSuperconductivityMagicangle2022},
this enhanced filling can arise because of both, $\H{layer}$ or $\H{Hartree}$.

To disentangle the effects of $\H{layer}$ and $\H{Hartree}$, we first consider the total filling required for $\numagic=3$ (taken here as a tentative upper bound for superconductivity) for TPG as a function
of the dielectric constants $\epsilon_\parallel$ and $\epsilon_\perp$, see  Fig.~\ref{fig:fig_2}c. To focus on the cascade physics, we
include 
moderate strain ($\strain =0.2\%$), which suppresses the appearance of correlated insulating states. 
For strong interactions (small dielectric constants), the entire $k=2$ nonmagic sector fills first before the magic sector starts to fill, incompatible
with the onset of superconductivity for $\nu\approx 2$ in Fig.~\ref{fig:fig_2}b.
In the opposite, weakly interacting limit, only negligible filling of the nonmagic sectors is induced, precluding an extended superconducting pocket.
Therefore, we use moderate $\epsilon_\parallel=14$ and $\epsilon_\perp=6$ in this section, referring to App.~\ref{appsec:extendeddata} for results for other parameter choices, including results at vanishing strain.

To probe the interplay of $\H{layer}$ and $\H{Hartree}$, Figs.~\ref{fig:fig_2}d,g show numerical results retaining only the in-plane Hartree and Fock terms (``HFX'') and
Figs.~\ref{fig:fig_2}e,h display corresponding results retaining only the out-of-plane ($\H{layer}$) and Fock terms (``XFL'').
Finally, Figs.~\ref{fig:fig_2}f,i include all terms (``HFL''). 
We first consider the band structures plotted in Figs.~\ref{fig:fig_2}d-f. Excluding the Hartree or layer potentials (HFX, Fig.~\ref{fig:fig_2}d and XFL, Fig.~\ref{fig:fig_2}e), we obtain only a minimal shift of the magic (red) vs.\ the nonmagic (blue) sectors. Interestingly, we find that in these approximations, the shifts due to $\H{Hartree}$ and $\H{layer}$ are largely compensated by the effects of $\H{Fock}$. However, there is a substantial shift when including all terms (HFL, Fig.~\ref{fig:fig_2}f). This highlights the importance of considering all of the terms together.

These trends are also reflected in the cascade plots in Figs.~\ref{fig:fig_2}g-i for $N=3,4,5$, which exhibit the flavor-resolved fillings as a function of $\nutotal$. Figures~\ref{fig:fig_2}g,h show results for XFL and HFX, respectively, and exhibit little 
effect of the nonmagic sectors on the cascade. This is consistent with the absence of a shift in Figs.~\ref{fig:fig_2}d and e. In contrast, Fig.~\ref{fig:fig_2}i shows increasingly delayed cascade transitions as the number of layers $N$ grows. This again reflects the importance of incorporating the effects of both, $\H{layer}$ and $\H{Hartree}$.

Numerically, for our choice of dielectric constants and 
$N=5$, the $\numagic=3$ cascade is pushed to $\nutotal\approx 5$,
while the $\numagic=2$ cascade happens at $\nutotal\approx3$. While the $\numagic=3$ cascade is consistent with experiment, the superconductivity data (Fig. \ref{fig:fig_2}b) suggest that the $\numagic=2$ cascade already appears at $\nutotal \approx 2$. Fully reproducing the experimental data may require more accurate modeling of the devices or more accurate approximations, e.g., allowing for the appearance of intervalley correlated ground states\cite{vafekKangStrongCouplingPhases2019,zaletelBultinckGroundStateHidden2020,bultinckKwanKekulSpiralOrder2021,bultinckParkerStraininducedQuantumPhase2021,yazdaniNuckollsQuantumTexturesManybody2023,nadj-pergeKimImagingIntervalleyCoherent2023}.

\section{Large-$N$ analysis}
\label{sec:largenanalysis}

We now consider the interplay of the in-plane Hartree, Fock, and layer potentials in the experimentally unexplored cases of $N > 5$
and $\k=2$
. The key question we would like to explore is to what extent TNG reproduces the phenomenology of TBG, when charge-inhomogeneity-induced band shifts are included? 

\begin{figure}[t]
  \includegraphics[width=0.5\textwidth]{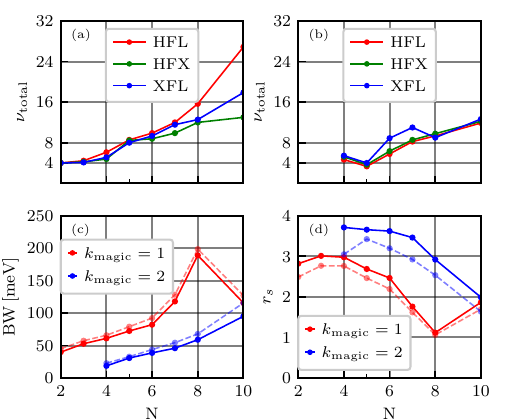} 
  \caption{(a) $\nu_{\text{total}}$ as a function of layer number $N$ at $\nu_{\text{magic}}=4$ choosing
  $k=1$ as the magic sector at $\epsilon_\parallel=10$, $\epsilon_\perp=6$, $\strain=0\%$.
(b) Same as (a) for $\k=2$.
(c) Bandwidth at $\nu_{\text{magic}}=4$ 
  for the choice of $\k=1$ (red) and $\k=2$ (blue). Dashed curves are for finite strain $\epsilon_{\text{strain}}=0.2\%$.
  (d) Effective interaction  parameter $r_s$ at $\nu_{\text{magic}}=4$ 
  for $\k=1$ (red) and $\k=2$ (blue). Dashed lines are at finite strain $\epsilon_{\text{strain}}=0.2\%$.}
\label{fig:figthree}
\end{figure}

Figure~\ref{fig:figthree} presents the main results of this section
for $\epsilon_\parallel=10$ and $\epsilon_\perp=6$. In Fig.~\ref{fig:figthree}a and Fig.~\ref{fig:figthree}b, we consider the $\nutotal$ needed to achieve complete filling of the magic sector, $\numagic=4$. We compare the cases of $\k=1$ (spectral weight peaked in the central layers, Fig.~\ref{fig:figthree}a) and $\k=2$ sector (spectral weight predominantly away from the central layers, Fig.~\ref{fig:figthree}b). Each figure shows plots including (i) the in-plane Hartree and Fock (HFX), (ii) the layer potentials and Fock (XFL), and (iii) all terms combined (HFL). For $\k=1$ (Fig.~\ref{fig:figthree}a), we see that the total filling required to completely fill the magic sector increases dramatically with $N$.
This confirms our expectation that gating the $\k=1$ sector becomes prohibitively difficult as the layer number increases.

Interestingly, when choosing $k=2$ as the magic sector (Fig.~\ref{fig:figthree}b), the magic sector fills much more easily. This is a result of the fact that the potential due to the gate charges is maximal at the central layers, so that the $k=1$ sector is more strongly shifted than the $k=2$ sector. As a result, $\k=2$ circumvents the electrostatic barrier present for gating the $k=1$ sector, providing a promising platform to study TBG-like physics in TNG samples with larger $N$.

In Fig.~\ref{fig:figthree}c, we consider the bandwidth of the magic sector. 
We compute the interacting bandwidth of the completely filled magic bands at $\numagic=4$ (see App.~\ref{appsec:extendeddata} for other choices) defined as
\def\bw{\max_{\v k} E^+_{\v k}- \min_{\v k}E^-_{\v k}}
\begin{equation}
\text{BW} =\bw.
\label{eq:bandwidth_def}
\end{equation}
Choosing $k=1$ (red) as the magic sector, we observe a substantial increase in bandwidth due to the in-plane Hartree and layer potentials. This suggests that even if the bands could be filled, the increased bandwidth will suppress correlated physics associated with the flat-band regime. 
Choosing $k=2$ (blue) as the magic sector, the bandwidth also increases with $N$, but less so than for $\k=1$.
This can be partially explained by the fact that much of the bandwidth is interaction driven and for a given $N$, $\k=1$ has a smaller unit cell than $\k=2$.
To accurately gauge the importance of interactions in the magic bands, we need to compare the bandwidth to the interaction scale. The effective interaction scale depends on the vertical spread of charges in the sector of interest. Using that the interaction between charge distributions with wave vector $\mathbf{q}$ in two layers separated by a distance $d$ is $(e^2/2\varepsilon_\parallel\epsilon_0 q) e^{-qd}$ (cf.~Eq.~\eqref{eq:appvbare}), the effective interaction energy per flat-band electron can be estimated as
\begin{align}
&\frac{e^2}{4\pi\epsilon_\parallel\epsilon_0 L_M}\left\langle\exp(-\lambda G|z-z'|)\right\rangle =\nonumber\\
&\quad =\frac{e^2}{4\pi \epsilon_\parallel \epsilon_0 L_M}\sum_{i,j} \Weight^{(k)}_{i}\exp(-\lambda Gd_l|i-j|) \Weight^{(k)}_{j}\,.
\end{align}
Here, the average in the first line is over the pairs of layers (located at $z$ and $z'$) accounting for the charge distribution of sector $k$ over layers as described by $W_i^{(k)}$. We also used that the characteristic wave-vector scale $G$ is given by the magnitude of the shortest reciprocal lattice vector $G=4\pi/(\sqrt{3}L_M)$, i.e., the
inverse of the  moir\'e length $L_M$. In the exponent, $\lambda$ accounts for the fact that the characteristic wave vector depends somewhat on the interaction effect of interest. We choose $\lambda=1$ for Hartree effects, and $\lambda = 0.5$ for correlation (Fock) effects. 

We can now use the computed bandwidth to define a dimensionless measure of the interaction strength in the flat bands,
\def\rs{e^2 \langle \exp{(-q|z-z'|)} \rangle /(4\pi\epsilon_\parallel \epsilon_0 L_M\text{BW})}
\begin{equation}
r_s = \rs.
\end{equation}
While this is still an oversimplified measure of interaction effects in flat bands \cite{bernevigSongMagicAngleTwistedBilayer2022}, it serves as a useful metric in comparison to similar analysis for TBG \cite{klugKlugChargeOrderMott2020}.
In Fig.~\ref{fig:figthree}d
we plot the effective $r_s$ as a function of layer number $N$. For $\k=1$ (red full line) and zero strain, $r_s$
decreases with increasing $N$, suggesting that devices with $N<5$ are most likely
to exhibit correlation effects. Strained $\k=1$ data (red dashed line) highlight the advantage of $N>2$. The importance of a given nominal value of strain diminishes with $N$. For this reason, $r_s$
is larger for strained $N=3$ than $N=2$.
Interestingly, we find that $r_s$ is larger for $\k=2$ (blue) than for $\k=1$. This holds even for strained devices. For increasing $N$, again, there is a decrease in $r_s$, which nevertheless stays above the $\k=1$ value.

To understand this peculiar behavior of $r_s$, we consider $N=5$ and $\k=2$ at zero strain. For $k=2$ at the magic angle, 
the $k=1$ sector is nominally  below the magic angle, but still very flat. This results in a cascade-like transition, at which 
the $k=2$ sector becomes almost completely filled, while  the $k=1$ sector has negative (hole) filling. 
Consequently, we find  $\nutotal<4$ at $\numagic=4$. 
This transition is encouraged by the central charge distribution over layers, larger inhomogeneity (see Sec.~\ref{app:spHartreetwistnagledep}), and larger bandwidth of the nonmagic, $k=1$ sector (with effective twist-angle below the magic angle).
After the cascade, the inhomogeneity of the holes from $k=1$ partially cancels against the inhomogeneity of the $k=2$ electrons, 
yielding a filled magic band with anomalously small Hartree broadening.

The behavior of $r_s$, together with the required doping dependence shown in Fig.~\ref{fig:figthree}a,b, suggest that to realize strongly interacting bands for large $N$ multilayer devices, it is necessary to focus on sectors $k\neq 1$ such that the spectral weight is not localized near the center of the device structure. For example, for the $k=2$ sector to be at the effective magic angle of $\thetaeff{2} = 1.1^\circ$, this would correspond to physical twist angles of $1.1^\circ$, $1.37^\circ$ for $N=5,6$-layer devices, respectively (see App.~\ref{sec:SM_Methods} for further analysis).

Finally, we comment on the role of dielectric constants in large-$N$ multilayers. In the literature, these constants are taken as fitting parameters
and frequently vary between experiments. Thus, it is helpful to discuss the
behavior of Fig.~\ref{fig:figthree} as a function of the dielectric constants. The
effect of a decreasing interaction strength on Fig.~\ref{fig:figthree}a is to shift all the curves downward (see Fig.~\ref{app_fig:straindepatmagiccharge}). At zero strain, changing $\epsilon_\parallel$ from $10$ to $14$ leaves the cascade physics unchanged, since it comes from two sets of very flat single-particle bands ($k=1$ and $k=2$).
At nonzero strain, decreasing interaction strength lowers $r_s$, as the strain-induced broadening becomes more relevant. Detailed parameter dependences are in App.~\ref{appsec:extendeddata}.

\section{Summary and Discussion}

In our analysis, we demonstrate how in-plane Hartree and layer potentials control the phase diagram of alternating-angle twisted multilayer graphene. Compared with the experimental results of Ref.~\cite{jarillo-herreroParkRobustSuperconductivityMagicangle2022,nadj-pergeZhangPromotionSuperconductivityMagicangle2022}, we showed that it is the interplay of these two effects that accounts for the filling enlargement of the superconducting pocket with layer number. In fact, we find that small-$N$ devices are the preferred layered structures to study $k=1$ flat-band physics. For $N > 5$, the magic sector present in the decoupling introduced in Ref.~\cite{vishwanathKhalafMagicAngleHierarchy2019} becomes strongly modified by the presence of Hartree effects to the extent that electrostatic doping of that sector becomes challenging.
In addition, the interacting bandwidth is enlarged by the in-plane and out-of-plane (layer) Hartree effects, likely precluding Fock-driven correlated phenomena.

The suppression of exchange-driven correlated phenomena by the Hartree effect relies on the mechanism of band shifting. Indeed this mechanism has been observed in the context of the TTG, where shifting of the flat band with respect to the Dirac cone can be seen spectroscopically \cite{nadj-pergeKimEvidenceUnconventionalSuperconductivity2022}. However, to date no scanning tunneling microscope (STM) experiments were carried out on $N>3$ devices. Such experiments may allow one to verify the scenario developed here.
This may also allow one to assess whether alternative theoretical explanations of the enlarged superconducting pocket, such as the more exotic scenarios discussed in Ref.~\cite{nadj-pergeZhangPromotionSuperconductivityMagicangle2022}, are necessary. We caution, however, that for  STM measurements, one side of the sample is typically left
uncovered, so that there is only one gate on the opposite side.
In this single-gate setup, it is impossible to vary displacement field and doping independently. Instead, varying gate voltage traces
out a line in the filling-displacement field plane. Nonetheless, we expect the qualitative physics of band shifting to persist as it is a robust consequence of charge inhomogeneity. However, quantitative predictions must be adapted to the new device geometry.

Experiments on moiré graphene systems exhibit substantial particle-hole asymmetry, unlike our theoretical analysis. Specifically, in TBG correlated insulators  appear to be more robust on the electron side than on the hole side. Similarly,  superconductivity can also appear in a particle-hole asymmetric manner \cite{nadj-pergePolskiHierarchySymmetryBreaking2022}. In the TPG samples studied in Ref.~\cite{nadj-pergeZhangPromotionSuperconductivityMagicangle2022}, superconductivity persists
up to $\nutotal = 5$ on the electron side,
but only down to $\nutotal=-4$ on the hole side. 
Particle-hole symmetry breaking can be incorporated into the BM model \cite{macdonaldXieWeakFieldHallResistivity2021,kaxirasCarrExactContinuumModel2019,vafekKangPseudomagneticFieldsParticlehole2023}. However we find this to be insufficient to reproduce the observed experimental trends. The presence of particle-hole symmetry is a common feature of theoretical efforts to date and requires further investigation.

While our results 
suggest that correlated phenomena are likely precluded for $N>5$ samples with $k=1$ magic sector, 
$k=2$ flat bands appear more promising. We find that $\k=2$ is subject to much weaker band reshuffling and thereby allows for effective electrostatic gating. Moreover, the $k=2$ band can become unexpectedly flat.
This suggests a resurgence of flat-band
physics for large $N$ in the $k=2$ sector, which could in principle differ from that seen
in TBG, for instance because  the multiple nearly flat bands may conspire to reduce the
Hartree-driven renormalizations that suppress the exchange effects.

\begin{acknowledgments}
We are grateful to Alex Thomson, Jason Alicea and \'Etienne Lantagne-Hurtubise for helpful discussions and collaboration on related projects.
Work at Freie Universit\"{a}t Berlin was supported by Deutsche Forschungsgemeinschaft through CRC 183 (project C02) and a joint ANR-DFG project (TWISTGRAPH). C.L. was supported by start-up funds from Florida State University and the National High Magnetic Field Laboratory. The National High Magnetic Field Laboratory is supported by the National Science Foundation through NSF/DMR-1644779 and the State of Florida. S.N-P acknowledges the support of NSF (award DMR-1753306) and the Office of Naval Research (award N142112635).   \end{acknowledgments}

\bibliography{zoteroexport}
\clearpage

\begin{widetext}
\appendix
\renewcommand\thefigure{\thesection\arabic{figure}}  
\setcounter{figure}{0}   
\renewcommand\thetable{\thesection\arabic{table}}   
\setcounter{table}{0}   

\section{Properties of the single-particle Hamiltonian}
\label{appsec:spphysics}
\subsection{Sector decomposition}
\label{appsubsec:spphysicssectordecomposition}
We review the  derivation of the sector decomposition, following Ref. \cite{vishwanathKhalafMagicAngleHierarchy2019}.
Labeling graphene layers by $i \in \{1,\ldots, N\}$, we have $n_e =\lfloor
N/2 \rfloor$ even layers with twist $\theta$ relative to the $n_o =\lceil N/2
\rceil$ odd layers.
Interlayer hopping only couples between odd and even layers. 
Thus, there can be a vector in layer space 
with support only in the odd layers which  
maps onto another vector with support only in the even layers under interlayer hopping.
This vector, in turn, maps back onto the first.
Mathematically, we are looking to find 
the singular-value decomposition (SVD) of 
the adjacency matrix $W$ in the space of the layers 
$(n_o,n_e)$.
This dimensionless matrix codifies between which layers there is hopping,
\begin{equation}
W =  \begin{pmatrix}
\begin{array}{cccc}
1 & 0 & 0 & \cdots \\ 1 &1 &0 & \\0&1&1& \\
    \vdots & & & \ddots 
\end{array}
    \end{pmatrix}\,,
\end{equation}
where $W_{i,j}=1$ if layers $2\cdot i-1$ and $2\cdot j$ are adjacent.
The SVD procedure yields right singular vectors $R_j^{(k)}$, left singular vectors $L_i^{(k)}$ and
eigenvalues $\Lambda_k$ satisfying $WR^k  =\Lambda_k L^k$. The eigenvalues
$\Lambda_k$ are the coefficients introduced in
Eq.\ \eqref{eq:thetaeffdef} rescaling the
interlayer hopping.
The $n_e$ dimensional right singular vector $R^{(k)}$ is the
wave function on physical, even layers  for the $k$-th twisted
bilayer graphene-like sector. Accordingly, the $n_o$ dimensional left singular
vector $L^{(k)}$ for $k\leq n_e$ gives the wavefunction across odd
physical layers for the $k$-th twisted bilayer graphene-like sector. For
$N$ odd, there is one additional left singular vector $L^{(n_o)}$, which spans
the kernel of $W^T$. This vector gives the spectral weight across the odd
layers of the MLG sector.
Further, in terms of the vectors $R^{(k)}$ and $L^{(k)}$, the basis transformation matrix $\Vtng$ is given by:
\begin{equation}
\label{eq:vtngdef}
\Vtng = 
 \begin{pmatrix}L^{(1)}_1 & 0 & L^{(2)}_1 & \cdots \\ 
 0 &R^{(1)}_1 &0 & \cdots \\
 L^{(1)}_2 &0&L^{(2)}_2& \cdots \\
    \vdots & \vdots & \vdots & \ddots \end{pmatrix}\,.
\end{equation}
The SVD procedure 
yields $n_e$ TBG-like sectors with hoppings renormalized by $\Lambda_k$ (and for $N$ odd, an extra Dirac cone corresponding to the kernel of $W^T$). 
We use the equation $WR^{(k)}  =\Lambda_k L^{(k)}$, together with its transpose, to obtain
$W^TWR^{(k)}=\left(\Lambda_k\right)^2 R^{(k)}$, which is a Hermitian eigenvalue problem.
We therefore need to find the eigenvalues and eigenvectors of the $n_e \times n_e$ symmetric matrix:
    \begin{equation}
    \label{eq:wtw}
        W^TW =  \begin{pmatrix}2 & 1 & 0 &0& \cdots \\ 1 &2 &1 &0 & \\0&1&2&1&  \\
    \vdots &  &  &  & \ddots \end{pmatrix},
    \end{equation}
where for $N$ odd $(W^TW)_{n_e,n_e} =1 \neq 2$.
This matrix can be physically interpreted as the Hamiltonian matrix 
of a tight-binding chain with open boundary conditions, on-site mass $2$ and
hopping of magnitude $1$.
To find the eigenvalues and eigenvectors, we start with solutions of
the infinite chain problem, which are
plane waves, $e^{i pj}$, for some momentum $p$, where
$j \in[-\infty,\infty]$ are the sites of the infinite chain. 
The physical sites of our open chain, corresponding
to the even layers, on the other hand,
go only from $1$ to $n_e$.
$e^{i pj}$ are eigenvectors of the infinite problem with
eigenvalue $2+2\cos(p)$. Note the degeneracy
$p \to -p$.
Specific combinations of these plane wave solutions for some
$p$ are in fact also solutions of the open chain. 
Due to the absence of next-nearest neighbor hopping, the 
only point at which the infinite solutions could fail to be solutions is
at the edges of the open chain. For example, take the $j=1$ boundary site.
$\exp(i pj)$ generally does not satisfy the open boundary problem,
as there is no hopping from the (non-existent) $j=0$ site
to $j=1$ in the open boundary problem.
However, taking a linear combination of $p$ and $-p$ 
to form $R_j = \sin (pj)$ has a zero at $j=0$, so in the infinite
problem the hopping from $j=0$ to $j=1$ does not contribute to
the equation. Therefore, $\sin(pj)$ are the class of wave functions that satisfy the open
boundary condition (BC) at the left end, $j=1$.
Now let us move to the boundary condition at $j=n_e$.
For $(W^TW)_{n_e,n_e}=2$ ($N$ odd), we simply need to require
that $\sin[p(n_e+1)]=0$, in order that the hopping
from the nonexistent $n_e+1$ site vanishes.
This leads to the quantization condition
$p(n_e+1) = k \pi$ with $k$ positive integer.

For $(W^TW)_{n_e,n_e}=2$, we need to analyze the equation
at site $n_e$.
The open BC equation reads:
\begin{equation}
R_{n_e-1}+R_{n_e} = E R_{n_e},
\end{equation}
while the periodic infinite solution satisfies the following:
\begin{equation}
R_{n_e-1}+2 R_{n_e} + R_{n_e+1} = E R_{n_e}.
\end{equation}
This suggests that if we find an infinite solution
with $R_{n_e} + R{n_e+1}=0$, it will also satisfy the open
boundary condition at $j=n_e$ with on-site lower mass.
To satisfy the left boundary condition, we need to have
$R_j \propto \sin(pj)$, 
so we have a condition on $p$:
$\sin(pn_e ) +\sin[p(n_e+1)]=0$.
This will be satisfied precisely when
$p(n_e+\frac{1}{2}) = k \pi$.
We can write the condition for $N$ odd and $N$ even as
one condition, using that for $N$ even, $n_e = N/2$ and
for $N$ odd $n_e = (N-1)/2$:
\begin{equation}
p(N+1) = 2\pi k.
\end{equation}
From this, the full solution for $R^{(k)}_j$ reads:
\begin{equation}
R^{(k)}_j =  \sqrt{\frac{4}{N+1}} \sin(2 \pi k j/(N+1)).
\end{equation}
The eigenvalues are 
\begin{equation}
E_k =  2 + 2 \cos \left [2\pi k/(N+1)\right] = 4 \cos^2 [\pi k/(N+1)],
\end{equation}
from which the singular values are (since $\Lambda_k^2 = E_k$):
\begin{equation}
\Lambda_k =  2  \cos \left [\pi k/(N+1)\right].
\end{equation}

We can also write down the 
$L^{(k)}_j$ using the condition for $k\leq n_e$
$WR^{(k)} = \Lambda_k L^{(k)}$,
while for $N$ odd there is an extra left singular vector $W^TL^{(n_o)} = 0$.
For $k\leq n_e$, we get:
\begin{equation}
L^{(k)}_j =  \sqrt{\frac{4}{N+1}} \sin[ \pi k (2j-1)/(N+1)].
\end{equation}
Lastly, for $N$ odd, we get an extra MLG-like sector with a vector
\begin{equation}
L^{(n_o)}_j = \frac{1}{\sqrt{n_o}} (-1)^j.
    \end{equation}
Having obtained the layer wavefunctions for each sector $k$, we consider the average
occupation of each layer for an electron in sector $k$.
Considering an electron in a TBG-like sector $k$ to be half in the odd layers and half in the even layers,
we obtain the density distribution across the layers
\begin{equation}
\label{eq:weightdistributionsectorsin}
\Weight^{(k)}_l \equiv
 \frac{1}{2}\left[(L^k_1)^2,(R^k_1)^2,(L^k_2)^2,\ldots,(\{L/R\}^k_{N})^2\right]_l
=\frac{2}{N+1} \sin^2\left[ \pi kl /(N+1)\right]
 \end{equation}
where the last entry in the definition is $L$ for $N$ odd and $R$ for $N$ even, 
obtaining Eq.~\eqref{eq:weightsevaluated}. The weights are plotted in Figure \ref{fig:fig_1}c.

\subsection{Twist angle dependence of the in-plane charge inhomogeneity} 
\label{app:spHartreetwistnagledep}
In Fig.~\ref{fig:twistdep}, we plot the 
dependence of the average wavefunction overlap
\begin{equation}
\overline{\braket{u_{\v{k+G},\alpha}|u_{\v k},\alpha}} =\frac{1}{N_{\v k}}\frac{1}{N_{\v G}} \sum_{\v{k},\v{G}}\frac{1}{2}\sum_{\alpha =1,2 }\left|\braket{u_{\v{k+G},\alpha}|u_{\v k,\alpha}}\right|
\end{equation}
on twist angle for $N=2$ (this result applies to any TBG-like sector) for the two central flat bands.
Here the sum over $\v G$ runs over the $N_{\v G}=6$ shortest nonzero reciprocal lattice vectors and
$\v k$ are in the first Brillouin zone, with $N_{\v k}=144$ the number of $\v k$ points in the numerical calculation grid.
This average overlap increases with decreasing twist angle. 
Its meaning can be understood from Eq.~\eqref{eq:hmfhartree}.
The larger the wavevector $\v{G}$ overlap, the more strongly a sector couples to the in-plane inhomogeneity at wavevector $\v{-G}$.
Converesely, sectors with a larger overlap at wavevector $\v{G}$ generate a larger mean-field inhomogeneity at $\v{-G}$.
This implies that for $\k=1$, the magic sector feels the in-plane potential most strongly and is most effective at generating it.

\begin{figure}[h]
  \includegraphics[width=1\textwidth]{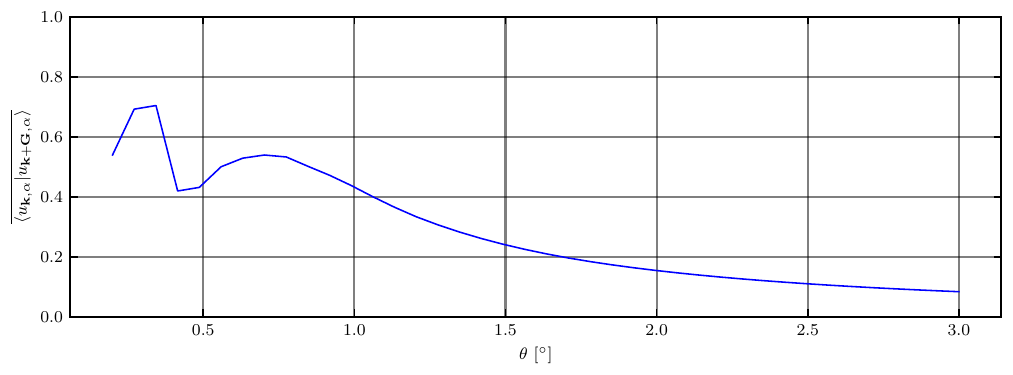}  
  \caption{
  Dependence of $\overline{\braket{u_{\v{k+G},\alpha}|u_{\v k},\alpha}}$, a quantity that controls the in-plane Hartree correction, on twist angle for $N=2$}
  \label{fig:twistdep}
\end{figure}
\subsection{Strain}
\label{app:strain}
In TBG, heterostrain drastically 
increases the single particle bandwidth \cite{fuBiDesigningFlatBands2019},
changes the nature of correlated states \cite{bultinckParkerStraininducedQuantumPhase2021,bultinckKwanKekulSpiralOrder2021}, and can induce in-gap states \cite{Kolar2023}.
The procedure for implementing heterostrain in TBG involves
adding vector potentials due to the changes in graphene hoppings and distorting the 
moiré Brillouin zone, altering the momentum space distance between the two layers of Dirac cones 
and the moiré reciprocal vectors.
Since, for a bilayer, any layer-dependent strain can be decomposed as the sum of hetero and homostrain, 
and homostrain has negligible effect, including heterostrain in this way 
is a generic procedure that captures qualitative physical trends.
In systems with more than two layers, there are more
nongeneric layer dependencies possible. 
As the purpose of our modeling is to introduce a mechanism for broadening the single-particle bandwidth,
we consider a simple procedure and only add the vector potentials induced by the graphene hoppings, choosing
an antisymmetric layer structure: 
\begin{equation}
\mathbf A_l = (-1)^l\mathbf A_0,
\end{equation}
where the single-layer vector potential is given by 
\begin{equation}
\mathbf A_0 =\frac{\sqrt{3}}{2a} \beta \left( \epsilon_{xx}-\epsilon_{yy},-2\epsilon_{xy}\right),
\end{equation}
with $a$ being the monolayer graphene lattice constant and $\beta \approx 3.12$ the hopping modulus factor
\cite{fuBiDesigningFlatBands2019}.
We choose $\epsilon_{xx} = \strain, \epsilon_{xy}=0$ and $\epsilon_{yy} = - 0.16\,\cdot \strain$ 
($0.16$ is the Poisson ratio for graphene),
varying $\strain$ from 
$0$ to $0.2 \cdot 10^{-2}$.
This layer structure is motivated by the fact that it acts just like a heterostrain vector
the potential within each bilayer-like sector at zero displacement field.
The above-defined vector potentials couple via minimal coupling to the momentum operator \cite{fuBiDesigningFlatBands2019}.
\subsection{Density of states for nonmagic sectors in the Dirac cone approximation}
\label{subsections:dos_non_int_dirac_cone}
In this section, we evaluate the numerical constants that appear in the expression for density of states (DOS) for a Dirac cone dispersion to obtain estimates for the DOS of the nonmagic sectors, as used in Section II of the main text. To this end, let us evaluate the prefactor of Eq.~\eqref{eq:formulanuk} with $v_D$ instead of $v_D^{(k)}$: 
\begin{equation}
\label{eq:evaluatedformulanuk}
\frac{A_{\text{uc}}}{4\pi (\hbar v_D)^2}  = 
\frac{\sqrt{3} (0.246 \si{nm})^2/(8\sin^2(\theta/2))  }{4\pi (6.582\cdot 10^{-16} \si{eV}\si{s} \cdot 10^6 \si{ms^{-1}})^2} =31.6/\theta^2 \si{eV^{-2}},
\end{equation}
where in the last equality, the twist angle $\theta$ should be plugged in degrees.
For the $k=2$ nonmagic sector in TPG, 
we have $N_f=4$, $c_{k=2}=2$,$\theta=1.9^\circ$, $v_D^{(k=2)}=0.35 v_D$.  We note in passing that 
Ref. \citenum{macdonaldBistritzerMoireBandsTwisted2011} finds a smaller Dirac velocity. This is because 
here we account for lattice corrugation by taking $w_{AA}/w_{AB} = \frac{8}{11}$, while in Ref.~\citenum{macdonaldBistritzerMoireBandsTwisted2011}
the unrelaxed value, $w_{AA}/w_{AB} = 1$, is taken.
Plugging into Eq.~\eqref{eq:evaluatedformulanuk}, we obtain
\begin{equation}
\nu^{TPG}_{k=2} = 
(5.71\cdot  10^{-4}\si{meV^{-2}}) \mu_2^2.
\end{equation}
As an example, for $\mu_2=10 \si{meV}$, using Eq.~\eqref{eq:formulanuk} we obtain filling
$\nu_2 \lesssim 0.06$.
As noted in the main text, for $\nu_2 \gtrsim 0.5$, we use the numerically computed full noninteracting density of states which involves a DOS peak at the
van Hove singularity.
\section{Interacting Hamiltonian}
\label{appsec:interacting}
In this section of the appendix, we discuss various elements of the analysis that were carried out in going from the full interacting Coulomb Hamiltonian for the 3D system
to the Hamiltonian, including only the layer indices. We also detail the
mean-field decoupling of the out-of-plane term.
\subsection{Integrating out the gate electrons}
\label{app:vbaretoveff}
Here we start from the full 3D Coulomb interaction $\frac{1}{2}\int d\v r \, d\v r' V(\v r-\v r') :\rho(\v r)\rho (\v r'):$
to obtain an effective interaction for TNG.
We consider the charges to be constrained in $N+2$ layers labeled
by an index $I$ going from $0$ to $N+1$ at vertical
positions $z_I$.
This corresponds to the physical situation of a sample
with $N$ graphene layers and two gate layers $I=0, N+1$.
In other words, we decompose 
$\rho(\v r) = \sum_I \rho_I(\v r) \delta(z-z_I),$
where $\rho_I(\v r)$ is the (2-dimensional) density in layer $I$.
In Fourier space, we have
\begin{equation}
\H{int}^{\text{bare}} =  \frac{1}{2A}\sum_{\v{q},I,J} \Vbare_{IJ}(\v q) :\rho_{I,\v q} \rho_{J,-\v q}:,
\end{equation}
where $A$ is the 2-dimensional area of the sample, we sum also over layers $0$ and $N+1$ corresponding to the gates, and
$\Vbare_{IJ}$ is the bare Fourier-transformed 2D 
Coulomb interaction with vertical separation 
$d_{IJ} = |z_I -z_J|$, which reads
\begin{equation}
\label{eq:appvbare}
\Vbare_{IJ}( \v q) = \frac{e^2}{2\epsilon \epsilon_0 q} \exp\left( - d_{IJ} q\right).
\end{equation}
For $\v q=0$, we separate the divergent and finite parts as follows
\begin{equation}
\label{eq:Vbareqtozero}
\Vbare_{IJ}( \v q \to 0) = \frac{e^2}{2\epsilon \epsilon_0 } \left[
O\left(\frac{1}{q}\right)  - d_{IJ} \right].
\end{equation}
The divergent part is canceled if the total charge adds up to zero
$\sum_I \rho_{I,\v q =0}  = 0$, and what remains of the $\v q =0$ term
is $-\frac{e^2}{2\epsilon \epsilon_0 } d_{IJ} $.
Therefore we obtain, separating $\v q =0$:
\begin{equation}
\H{int}^{\text{bare}} =  \frac{1}{2A}\left[\sum_{\v{q}\neq 0,I,J} \Vbare_{IJ}(\v q) :\rho_{I,\v q} \rho_{J,-\v q}:
- \sum_{I,J} \frac{e^2}{2\epsilon \epsilon_0 } d_{IJ} :\rho_{I,\v q =0} \rho_{J,\v q=0}:
\right].
\end{equation}
which still includes the gate charges.
We can simplify the second term by working at the fixed gate and sample charge,
allowing us to replace
$\frac{\rho_{0,0}}{A} = -\frac{n}{2}$
$\frac{\rho_{N+1,0}}{A} = -\frac{n}{2}$, and
$\frac{\sum_{i=1}^{N} \rho_{i,0}}{A} = n$. 
Then it can be (up to a $n$ dependent constant) more physically rewritten as the 
electrostatic energy of the perpendicular electric field
between the layers, which is given by Gauss' law as: 
\begin{equation}
E^\perp_{i,i+1} = -\frac{e}{\epsilon_0 \epsilon}\left\{
\frac{1}{A} \sum_{l=1}^i \langle \rhoproj{l,0}\rangle-\frac{n}{2} \right\}.
\end{equation}
With this identification, $\H{int}^{\text{bare}}$ reads:
\begin{equation}
\H{int}^{\text{bare}} =  \frac{1}{2A}\left[\sum_{\v{q}\neq 0,I,J} \Vbare_{IJ}(\v q) :\rho_{I,\v q} \rho_{J,-\v q}: 
+\sum_{i=1}^{N-1} \epsilon\epsilon_0 d_l\frac{(E^{\perp \,}_{i,i+1})^2}{2} \right]
\end{equation}
For the $\v q \neq 0 $ term, we integrate out the gate electrons and
end up with an effective screened interaction for the N layers, 
whose form is obtained in the next section using the method of images.
Above, it was assumed that there is 
a single dielectric constant for the medium 
between the graphene layers and between the sample and the gates.
Here we consider the more realistic possibility of 
having different dielectric constants in between the graphene
layers and around the gates.
This leads to two modifications in 
Eq.~\eqref{eq:appinteracting_hamiltonian}:
Firstly, the perpendicular electric field term should 
have its own dielectric constant $\epsilon_\perp$, related to the out-of-plane dielectric properties of graphene.
Secondly, $V_{ij}(\v q)$ has a more complicated dependence
than in Eq.~\eqref{eq:appvbare}, since interaction at different scales sees different dielectric environments.  
We will include the first effect,
but for the sake of simplicity, we will model
$V_{ij}(\v q)$ as if there was a single dielectric constant, 
deriving its form in Section \ref{app:layerdepcoulombV} below.
However, we will allow the dielectric constant of
$V_{ij}(\v q)$ ($\epsilon_\parallel$) to differ
from $\epsilon_\perp$.
With this we obtain the effective system interaction Hamiltonian from the main text:
\begin{equation}
\label{eq:appinteracting_hamiltonian}
\H{int} =  \frac{1}{2A}\sum_{\v{q}\neq 0,i,j} V_{ij}(\v q) :\rho_{i,\v q} \rho_{j,-\v q}:
+\sum_{i=1}^{N-1} A \epsilon_\perp\epsilon_0 d_l\frac{(E^{\perp \,}_{i,i+1})^2}{2}.
\end{equation}

\subsection{Layer-dependent in-plane Coulomb interaction}
\label{app:layerdepcoulombV}
The interaction between two electrons depends on which layer each electron is in. 
In free space, this simply adds a factor $e^{-q|z-z_0|}$ in 
the Fourier transform of the interaction.
Here we calculate the layer-dependent interaction in Fourier space
in the presence of two gates at positions $z=\pm d_s$,
where $d_s$ is the screening length. We use the method of 
images, which solves the Poisson equation in the region $z in (-d_s,d_s)$ with the boundary condition $\partial_\perp V|_{\pm d_s}=0$ by
placing image charges above and below the gates.
First, we consider the positions of image charges 
when a positive unit charge is placed at $z_0$.
Due to the presence of two gates, there will be infinitely many 
image charges in the regions above $d_s$ and below $-d_s$.
We denote the z-coordinate of the position of the $n$-th image 
charge in the top gate ($z>d_s$) as $\dtop{n}$, while the z-coordinate 
of the position of the $m$-th image charge in the bottom gate will
be $\dbot{m}$.
The first image charge in the top gate will be at 
$\dtop{1}=2d_s-z_0$, while the first image charge in the bottom gate 
at $\dbot{1}=-2d_s-z_0$, and they have negative unit charge.
Next, the bottom gate is affected by the image charge 
in the top gate and vice versa, implying we need to place more and more charges.
We, therefore, obtain the intertwined recurrence relation for the positions of the $n+1$-th image charges
\begin{eqnarray}
\dtop{n+1} = 2d_s-\dbot{n} \\
\dbot{n+1} = -2d_s-\dtop{n},
\end{eqnarray}
where the charge of the $n$-th charge is $(-1)^n$.
This recurrence is solved by :
\begin{eqnarray}
\dtop{n} = 2nd_s+(-1)^nz_0 \\
\dbot{n} = -2nd_s+(-1)^nz_0.
\end{eqnarray}
The potential at vertical position $z$ and an 
in-plane distance $r$ away from the 
unit test charge is given by the sum of 
the potentials of the charge and all the image charges generated.
We have 
\begin{equation}
V(r,z,z_0) = \frac{1}{4\pi \epsilon \epsilon_0} \left[
\frac{1}{\sqrt{r^2+(z-z_0)^2}}
+\sum_{j=1}^{\infty}  
\frac{(-1)^j}{\sqrt{r^2+(2jd_s+(-1)^jz_0 - z)^2}} +
\frac{(-1)^j}{\sqrt{r^2+(2jd_s+z-(-1)^jz_0)^2}}
\right].
\end{equation}
In Fourier space, we obtain:
\begin{equation}
V(q,z,z_0) = \frac{1}{2\epsilon \epsilon_0}\frac{1}{q} \left\{
\exp(-q|z-z_0|)
+\sum_{j=1}^{\infty}  
(-1)^j \exp[-q(2jd_s+(-1)^jz_0 - z)]+ 
(-1)^j \exp[-q(2jd_s-(-1)^jz_0 + z)],
\right\}
\end{equation}
where we removed the absolute value in the image charge potentials since we 
are interested in the potential inside the sample, assuming $|z|<d_s, |z_0|<d_s$.
The sum over $j$ can be easily performed by separating
into $j$ odd and even, leading to the result:
\begin{align} 
 V(q,z,z_0)  && =
&&\frac{1}{2\epsilon \epsilon_0}\frac{1}{q}\cdot \left(\frac{e^{-q (z+z_0)} 
\left(-e^{2 q (d+z+z_0)}-e^{2 d q}+e^{2 q z}+e^{2 q z_0}\right)}{e^{4 d q}-1}+e^{-q |z-z_0|} \right).
\end{align} 
For $z=z_0=0$, $V(q,z,z_0) $ reduces to the $\tanh(qd_s)/q$ form usually used for double-gate screened interaction. 
On the other hand, with no screening ($d_s \to \infty$) we recover the bare
interaction in Eq.~\eqref{eq:appvbare}.

\subsection{Mean-field decoupling of out-of-plane electric field term}
\label{app:hartreefockderivation}
Here we detail the mean-field decoupling 
the out-of-plane ($\v q=0$) term.
For notational simplicity, we work out the general form before
projecting onto a fixed number of active
bands. We perform the mean-field decoupling of $\H{int}^{(\v q=0)}$:
\begin{equation}
\H{int}^{(\v q=0)}=
-\frac{1}{2A}\sum_{I,J} \frac{e^2}{2\epsilon_\perp \epsilon_0 } d_{IJ} 
\rho_{I,\v q =0} \rho_{J,\v q=0} \,\,\,
= 
\sum_{i=1}^{N-1} A \epsilon_\perp\epsilon_0 d_l\frac{(E^{\perp \,}_{i,i+1})^2}{2}
 + \,\, \text{Const\,,}
\end{equation}
which was derived assuming a fixed amount of charge on the gates, but still includes it explicitly (by summing $I,J$ from $0$ to $N+1$). 
We dropped the normal ordering
symbol since it only matters for $I=J$, for which
the vertical distance $d_{IJ}$ vanishes.
Let us recall the three constraints
\begin{itemize}
\item $\rho_{0,0}=-A \frac{n}{2}$
\item $\rho_{N+1,0} = -A\frac{n}{2}$
\item $\sum_{i=1}^{N} \rho_{i,0} = An$
\end{itemize}
For the mean-field decoupling, we use the $\v q=0$ layer density
form of the interaction.
Following standard procedures, there will be the Hartree
term, which corresponds to classical electrostatics
\begin{equation}
\H{layer}^{\text{Hartree}} =
-\sum_{I\neq J} \frac{e^2}{2\epsilon_\perp \epsilon_0 A} d_{IJ} 
\rho_{I,\v q =0} \langle \rho_{J,\v q=0} \rangle
=
\sum_{i=1}^N {\rho_{i,0}} (-e V_i)\,,
\end{equation}
where we changed sum over $I$ (from $0$ to $N+1$, including gates) 
to a sum over $i$ (from $1$ to $N$) since the gates have a fixed
charge.
Therefore the potentials are given by
\begin{equation}
V_i =   \frac{e}{2\epsilon_\perp \epsilon_0 A} \sum_J d_{iJ} \langle \rho_{J,\v q=0} \rangle.
\end{equation}
It is insightful to consider the potential difference 
between two neighboring layers
\begin{equation}
\label{eq:potdifffirstshot}
V_{i+1} - V_{i} =\frac{e}{2\epsilon_\perp \epsilon_0 A } \sum_J (d_{i+1,J} -d_{i,J}
) \langle \rho_{J,\v q=0} \rangle\,,
\end{equation}
where
\begin{equation}
d_{i+1,J} -d_{i,J} = 
\begin{cases}
d_l & \text{for } i \geq J\\
-d_l & \text{for } i < J .
\end{cases}
\end{equation}
With this relation, we can rewrite
Eq.\ \eqref{eq:potdifffirstshot}
\begin{equation}
\label{eq:second shot}
V_{i+1} - V_{i} =d_l \frac{e}{2\epsilon_\perp \epsilon_0 A } 
\left[\sum_{J\leq i} \langle \rho_{J,\v q=0} \rangle
-\sum_{J> i} \langle \rho_{J,\v q=0} \rangle \right].
\end{equation}
Since $\rho_0 = \rho_{N+1} = -n A/2$, the gate charge terms cancel. 
Further, since the total charge on the sample is fixed, we also have
\begin{equation}
-\sum_{i<J\leq N} \langle \rho_{J,\v q=0} \rangle =
\sum_{1 \leq J\leq  i} \langle \rho_{J,\v q=0} \rangle - nA .
\end{equation}
which yields
\begin{equation}
\label{eq:potdiffthirdandlastshot}
V_{i+1} - V_{i} =
d_l \frac{e}{\epsilon_0 \epsilon}\left\{
\frac{1}{A} \sum_{l=1}^i \langle \rhoproj{l,0}\rangle-\frac{n}{2} \right\}
= -d_l E^\perp_{i,i+1}.
\end{equation}
In the above expression, we identified
that the interlayer electric field is given by Gauss' law, Eq.~\eqref{eq:gausslaw}.

Next we consider the $\v q=0$ Fock term.
As the Fock term involves an integral over a range momenta
and is intensive, if we fix a single momentum term $\v q =0$ 
(as we do for the interlayer potential term), it will vanish in the thermodynamic limit.
Therefore we only need to keep the $\v q=0$ Hartree term.
Finally, in our numerics, we project on
a finite number of bands\, replacing $\rho_{l,0}$ by $\rhoproj{l,0}$.
 
\section{Analytical results on the layer potentials}
\label{appsec:analyticalunderstanding}
\subsection{Layer potentials in sector basis}
\label{app:layerpotentialsinsectorbasis}
In this section, we consider the mean-field layer potential term,
and rewrite it in the sector basis. We use the unprojected
form of the layer Hamiltonian:
\begin{equation}
\H{layer}^{\text{unprojected}} = \sum_l {\rho_{l,0}} (-e V_l),
\end{equation}
but the conclusions will also hold after projection.
To proceed,  we need to write $\rho_{l,0}= 
 \sum_{f,\v k,z}d^\dagger_{f,l, \v k,z} d_{f, l,\v k,z}$,
where $d^\dagger_{f,l, \v k,z}$ creates an electron in flavor $f$,
layer $l$, momentum $\v k$ and a joint sublattice/spin index $z$.
Since the transformation into sectors does not affect flavor, momentum,
or sublattice and spin, we will in the following omit their labels.
Using the SVD procedure, we can go from layer basis to
sector basis using the unitary basis transformation $\Vtng$ as
follows:
\begin{equation}
f^\dagger_s  =  \sum_l d^\dagger_{l}\Vtngind{ls} ,
\end{equation}
where $f^\dagger_s$, $s \in \{1,\dots, N \}$ creates an electron in the effective layer
$s$, which can either have support in the odd physical layers or even. 
As shown in Eq.~\eqref{eq:vtngdef}, 
the orthogonal matrix $\Vtng$ is 
closely related to the singular vectors $R^{(k)}, L^{(k)}$.
We therefore rewrite
\begin{equation}
 \sum_l {\rho_{l,0}} (-e V_l)  = \sum_{s,s'}
f^\dagger_s f_{s'}\sum_l \Vtngind{ls}\Vtngind{ls'} (-e V_l).
\end{equation}
To emphasize the sector (recall for $N$ layers there are $\lceil N/2 \rceil$ sectors labeled by index $k$) diagonal and off-diagonal
terms, we now switch $s$ for
a multi-index $k,i$, where $k \in \{ 1, \ldots, \lceil N/2 \rceil \}$ labels
the sector, and $i$ labels the effective odd or even layer of that
sector. For an MLG-like sector, this index is trivial.
With this rewriting, we write suggestively
\begin{equation}
\label{eq:Hlayersectordecomp}
 \sum_l {\rho_{l,0}} (-e V_l)  = \sum_{k,i}
f^\dagger_{k,i} f_{k,i}\sum_l \Vtngind{l,ki}\Vtngind{l,ki} (-e V_l)
+ \sum_{k \neq k',i} 
f^\dagger_{k,i} f_{k',i}\sum_l \Vtngind{l,ki}\Vtngind{l,k'i} (-e V_l),
\end{equation}
where we used the fact that $\Vtngind{l,ki}\Vtngind{l,k'i'}\propto \delta_{i,i'}$,
so that there are no layer index ($i,i'$) off-diagonal terms.
On the other hand, odd and even layer index preserving terms
are allowed. 
\subsubsection{Sector diagonal terms}
\label{sec:appsectordiagonalterms}
In this section, we focus on the sector diagonal terms,
which correspond to the first term in the Equation~\eqref{eq:Hlayersectordecomp}.  
For a TBG-like sector $k$, this term 
is a potential
$V_1 = \sum_l \Vtngind{l,k1}\Vtngind{l,k1} (-e V_l)$ 
on the effective odd layer and
$V_2 = \sum_l \Vtngind{l,k2}\Vtngind{l,k2}(-e V_l)$
on the effective even layer.
Decomposing the effective layer potential matrix $\begin{pmatrix} 
V_1 & 0 \\
0 & V_2 
\end{pmatrix}$ 
into layer-even and layer-odd components, we obtain that the 
effect of layer potentials within a sector
is twofold. 
It causes a shift of the whole sector by 
$U_k =  \frac{V_1 +V_2}{2} $
and an interlayer potential difference
$D_k =  V_1 -V_2 $ between the effective odd and even layers.
We can obtain an analytical formula for the
sector shift in terms of the matrix $\Vtngind{l,ki}$ and
therefore also in terms of the vectors $R^{(k)}, L^{(k)}$. 
\begin{equation}
\label{eq:firstformulaforUk_app}
U_k = \half \sum_{l,i} \Vtngind{l,ki}\Vtngind{l,ki}(-e V_l).
\end{equation}
In the above, we identify
\begin{equation}
 \frac{1}{2} \sum_{i} \Vtngind{l,ki} \Vtngind{l,ki}=\frac{1}{2}\left[(L^k_1)^2,(R^k_1)^2,\ldots,(\{L/R\}^k_{N})^2\right]_l =\Weight^{(k)}_l
\end{equation}
as the layer distribution weights $\Weight^{(k)}_l$, plotted in Figure \ref{fig:fig_1}c.
The final formula for the shift of the sector $U_k$ therefore reads:
\begin{equation}
\label{eq:formulaforUk_app}
U_k = \sum_l W^{(k)}_l (-e V_l).
\end{equation}
The derivaation of the interlayer potential difference
proceeds analogously, 
so we only give the expression,
which differs by an
extra $(-1)^l$ in the sum over layers
\begin{equation}
\label{eq:formulaforDk}
D_k = 2 \sum_l (-1)^l\, W^{(k)}_l (-e V_l).
\end{equation}
This $(-1)^l$ leads to a cancellation when compared to $U_k$. 
\subsubsection{Sector off-diagonal terms}
We now turn to the sector mixing terms, which
correspond to the $k\neq k'$ term in Eq.~\eqref{eq:Hlayersectordecomp}. 
Given that the potential difference between layers can become quite sizeable for large dopings, sector mixing will become important for large $N$.
If sector mixing is small, one can directly relate the physics to the TBG physics. On the other hand, for
large sector mixing, such direct mapping is no longer possible, and the bands become rather different from bare TBG-like bands.
However, these bands may still favor superconductivity and strong correlation physics, as seen in TTG under a displacement field.
One advantage arises for $N$ odd. In that case, opposite mirror symmetry eigenvalues forbid
mixing between adjacent sectors ($k$ and $k+1$, say).

\subsection{Evaluation of sector shifts}
\label{app:analyticlayerpotentials}
Given the layer structure of the sectors, we can obtain an
mean-field layer Hartree shift $\Delta U_k$ analytically.
We start with the layer vectors for sector $k$, obtained from the singular value decomposition from Sec.~\ref{appsubsec:spphysicssectordecomposition}.
For a general TBG-like sector, this corresponds to two vectors, 
$L^{(k)}$ giving the wave function of the effective odd layer 
across the odd physical layers, and 
$R^{(k)}$ giving the wavefunction of the effective even layer
across the even physical layers.
Using the results derived above in Section~\ref{sec:appsectordiagonalterms},
we can obtain the sector shift $U_k$ in terms of the weights 
$\Weight_{l}^{(k)}$ and the layer potentials $V_l$.
We obtain the layer potentials by using that a sector with filling $\nu_k$ 
has on average the following layer number density distribution
\begin{equation}
\langle \hat\rho_{l,0}\rangle =\frac{1}{\Auc} \Weight_{l}^{(k)} \nu_k.
\end{equation}
Knowing this, and using Eq.~\eqref{eq:gausslaw} the
electric field between two layers caused by sector filling $\nu_k$
(which causes an electron density $e\nu_k/(2\Auc)$ on the gates) becomes
\begin{equation}
E^\perp_{i,i+1} = -e \frac{-1/2 +  \sum_{l=1}^i \Weight^{(k)}_l   }{A_{\text{UC}}\epsilon_0 \epsilon_\perp}\nu_k.
\end{equation}
Using the formula 
Eq.~\eqref{eq:weightdistributionsectorsin} for $\Weight^{(k)}_l$,
we evaluate the sum of the weights
\begin{equation}
\sum_{l=1}^i \Weight^{(k)}_l = \frac{1}{N+1}\left[ i +1/2 - \frac{\sin\left[\pi k (2i+1)/(N+1)\right]}{2 \sin\left[\pi k/(N+1)\right]}\right].
\end{equation}
As a check, for $i=N$, we obtain $\sum_{l=1}^N \Weight^{(k)}_l = 1$, while for $N$ even, $i=N/2$, we get
$\sum_{l=1}^N \Weight^{(k)}_l = 1/2$, so that $E_{i,i+1}=0$ in the middle spacing.
Using that $V_{l+1}-V_{l}=-d_l E_{l,l+1}$, 
we can now integrate the electric field to calculate the
electron energy shift $-eV^{(k)}_l$ in layer $l$ due to the
filling of sector $k$:
\begin{equation}
\label{eq:vlayeranalytic}
-e V^{(k)}_{l+1} =  \nu_k  \frac{e^2 d_l }{\epsilon_0 \epsilon_\perp}
\left\{
l \cdot \left[ \frac{N-l-1}{2(N+1)}\right] + \frac{\cos\left[2 \pi k/(N+1)\right]-\cos\left[2 \pi k(l+1)/(N+1)\right]}
{4(N+1) \sin^2\left[\pi k /(N+1)\right]}
\right\}.
\end{equation}
We note that the maximal potential magnitude is in the middle of the sample, which is intuitive,
given that charge of a single sign is being distributed across the layers.

Having obtained the layer shifts due to the filling of
a single sector $k$,
we can now add the contributions due to all the sectors and
obtain $-e V_l$.
Using this, we get the sector shifts $U_k$, and
therefore also the numerical coefficients $\C{k,k'}$
giving the shifts of sectors in terms of the sector fillings
\begin{equation}
U_k =\sum_l \Weight^{(k)}_l (-e V_{l}) = \sum_{lk'} \Weight^{(k)}_l   
\nu_{k'}  \frac{e^2 d_l }{\epsilon_0 \epsilon_\perp}
\left\{
l \cdot \left[ \frac{N-l-1}{2(N+1)}\right] + \frac{\cos\left[2 \pi k'/(N+1)\right]-\cos\left[2 \pi k'(l+1)/(N+1)\right]}
{4(N+1) \sin^2\left[\pi k' /(N+1)\right]}
\right\}.
\end{equation}

Recalling the definition of $\C{k,k'}$ from 
Equation~\eqref{eq:capacitanceformula},
we can identify $\C{k,k'}$ as
\begin{equation}
\label{eq:ckkprimeformula}
\C{k,k'} = 
\sum_l
W^{(k)}_l
\left\{
l \cdot \left[ \frac{N-l-1}{2(N+1)}\right] + \frac{\cos\left[2 \pi k'/(N+1)\right]-\cos\left[2 \pi k'(l+1)/(N+1)\right]}
{4(N+1) \sin^2\left[\pi k' /(N+1)\right]}
\right\}.
\end{equation}
This equation is used to generate the Table \ref{tab:ckkprime}
in the main text for $N=4,5$.
At fixed $k,k'$, but taking $N \to \infty$, 
we can obtain $\C{k,k'}$ analytically by going from a sum 
to an integral in Eq.~\eqref{eq:ckkprimeformula}.
This immediately reveals a scaling with $N$.
We get for the dominant $O(N)$ terms:
\begin{equation}
\C{k,k'}=N \int_0^1 dy \sin^2(\pi k y) 
\left\{ y(1-y) + \frac{1-\cos(2\pi k' y)}
{2\pi^2 (k')^2}\right \}.
\end{equation}
Note that the integral over $y$ depends only on $k$ and $k'$,
with the entire $N$ dependence factored out in the front.
Evaluating this integral for $k,k' =1,2$, we obtain the large
$N$ entry in Table \ref{tab:ckkprime}.
In Table \ref{suptab:ckkprime}, we give the results for $\C{k,k'}$ in expression form, rather
than evaluated numerically as in the main text.
\begin{table}
\begin{tabular}{c|c|c|c}
$N$ & $\C{1,1}$ & $\C{1,2}$ = $\C{2,1}$ &$\C{2,2}$ \\
\hline
$4$  & $2\phi^4/(2+2\phi^2)^2$  & $2\phi^2/(2+2\phi^2)^2$ &  $2/(2+2\phi^2)^2$\\
\hline
$5$  & $29/72$ & $15/72$  & $9/72$  \\ 
\hline
$N \to \infty$  & $N (1/12+5/(8\pi^2))$  &$N (1/12+5/(16\pi^2))$ &
$N (1/12+5/(32\pi^2))$
\end{tabular}
\caption{Inverse capacitance $\C{k,k'}$ for $N=4,5$ layers and large $N$ for $k,k' \in \{1,2\}$ in expression form.  }
\label{suptab:ckkprime}.
\end{table}

For reference, we evaluate the numerical constants:
\begin{equation}
\label{eq:supp:evaluated}
\frac{e^2d_l}{A_{\rm{uc}}\epsilon_0} = \frac{e^2 \cdot 0.3 \si{nm}}
{\frac{\sqrt{3}\cdot  0.243^2}{2 (\pi/180)^2 \theta^2}\si{nm^2} \cdot e^2 \cdot 55.263 \, \si{  keV^{-1}nm^{-1}}} =  32.34 \theta_{\text{physical}}^2 \si{meV},
\end{equation}
with $\theta$ in degrees and
where we used
vacuum permittivity $\epsilon_0 = 55.263 \, \si{ e^2 keV^{-1}nm^{-1}}$ and
interlayer distance $d_l = 0.3 \si{nm}.$ This yields 
\begin{equation}
U_{k} =  \left[32.34 \frac{\theta_{\text{physical}}^2}{\epsilon_\perp} 
\sum_{k'}^{n_o} \C{k,k'} \nu_{k'} \right] \si{meV}\,.
\end{equation}

\subsection{Application to TPG}
For example, the $k=2$ sector in TPG has the 
following singular vectors
$$L_j^{(k=2)} = \frac{1}{\sqrt{2}}(1,0,-1)_j,R^{k=2}_j =  \frac{1}{\sqrt{2}} (1,-1)_j.$$
The weigths of the $k=1,2,3$ sectors are
\begin{eqnarray}
\Weight_l^{(k=1)} = \frac{1}{12}\left(1,3,4,3,1\right)_l \\
\Weight_l^{(k=2)} = \frac{1}{4}\left(1,1,0,1,1\right)_l \\
\Weight_l^{(k=3)} = \frac{1}{3}\left(1,0,1,0,1\right)_l. 
\end{eqnarray}
Evaluating, using $\theta = 1.9^\circ$,
interlayer $\epsilon_\perp \in [2,12]$,
the nonmagic effective chemical potential increases by
\begin{equation}
U_1-U_2  = 3.24/\epsilon_\perp \left[3 \nu_2 + 
7 \numagic \right] \si{meV}.
\end{equation}
Supposing that $\numagic= 4$, 
we obtain a range of
$ \Delta U \approx 7-\SI{45}{meV}$ increase
of the effective nonmagic sector chemical potential due to 
Hartree layer potentials.

We now consider effects of the layer potentials beyond simple sector shifts, which are:
\begin{itemize}
\item Intrasector potential difference, both for $k=1$ and $k=2$
\item A term mixing $k=1$ and $k=3$ -- magic and MLG-like, acting like an external displacement field in TTG
\end{itemize}
We can readily evaluate the magnitudes of all these terms assuming fixed sector filling using the results from the previous section.
We evaluate $-eV_l$ in terms of $\nu_1$ ($\numagic$), $\nu_2$:
\begin{equation}
-eV_l = \frac{e^2 d_l }{\epsilon_0 \epsilon_\perp \Auc}\left[\nu_2 \left(0,\frac{1}{4},\frac{1}{4},\frac{1}{4},0\right)_l + 
\nu_1 \left(0,\frac{5}{12},\frac{7}{12},\frac{5}{12},0\right)_l \right].
\end{equation}
With this in hand, we can evaluate:
\begin{equation}
D_1 = -\frac{e^2 d_l }{\epsilon_0 \epsilon_\perp \Auc}\left[\frac{1}{12}\nu_2 +\frac{1}{36}\nu_1\right],
\end{equation}
for the magic sector and:
\begin{equation}
D_2 = -\frac{e^2 d_l }{\epsilon_0 \epsilon_\perp \Auc}\left[\frac{1}{4}\nu_2 +\frac{5}{12}\nu_1\right],
\end{equation}
for the nonmagic TBG-like sector, singificantly larger than $D_1$. 
By mirror symmetry, the $k=2$ sector doesn't mix any other sector.
Let us however evaluate the mixing term of $k=1$ and $k=3$.
This is the term:
\begin{equation}
H_{13}= f^\dagger_{k=1,i=1} f_{k'=3,i=1}\sum_l \Vtngind{l,k=1,i=1}\Vtngind{l,k'=3,i=1} (-e V_l) + \text{h. c.}
\end{equation}
from Equation.~\eqref{eq:Hlayersectordecomp}, which we readily evaluate using $-eV_l$: 
\begin{equation}
H_{13}=-\frac{e^2 d_l }{\epsilon_0 \epsilon_\perp \Auc}\left[\frac{\sqrt{3}}{8}\nu_2 +\frac{7\sqrt{3}}{24}\nu_1  \right]f^\dagger_{k=1,i=1} f_{k'=3,i=1}  + \text{h. c.}
\end{equation}
$H_{13}$ has exactly the same effect as a displacement field in TTG. However, rather than being explicitly tunable in a doubly-gated setup, 
it is self-generated and doping dependent.

\section{Extended data }
\label{appsec:extendeddata}
\setcounter{figure}{0}  
\subsection{Extended data for $N=3,4,5$ }
\label{appsec:extendeddatatqgtpg}
This section presents extended data for $N=3,4,5$ as a function of various model parameters. 
In Fig.~\ref{fig:app:unstrainedtwofourfive}, we show the flavor resolved magic sector filling dependence on $\nutotal$ 
for $N=3,4,5$ at zero strain. 
The trends are qualitatively similar to the ones seen for finite strain. 
However, due to the constant density of states above the correlation induced gap, flavor polarization is preferred 
already upon infinitesimal doping from charge neutrality.
Further, compared to $\strain=0.2\%$, the $\numagic=3$ cascade appears earlier for HFL.
\begin{figure}[h]
  \includegraphics[width=1\textwidth]{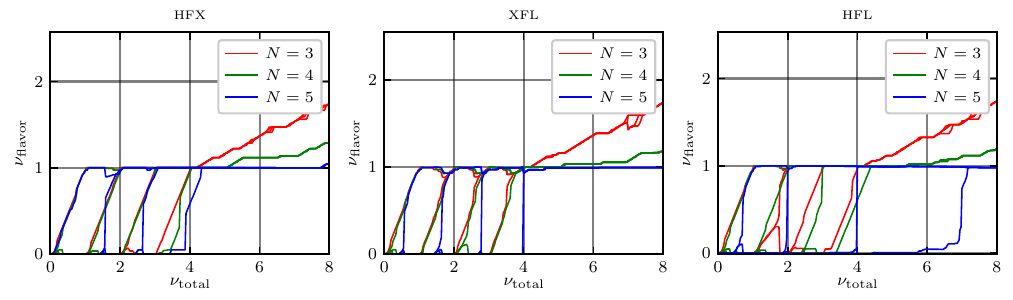}  
  \caption{Same as Fig.~\ref{fig:fig_2}d-f in
  the main text but without imposed strain. }
\label{fig:app:unstrainedtwofourfive}
\end{figure}

In Fig.~\ref{app_fig:strainedepsilondependencetpg}, we consider (as in the main text) a finite strain $\strain= 0.2\%$
at different values of the interaction strength parameters $\epsilon_\perp$ and $\epsilon_\parallel$ for $N=5$. 
To compare differeent interaction strengths most clearly, we plot the total filling of the magic sector $\numagic$ rather
than flavor resolved fillings. 
As argued in the main text, we find that the stronger the interaction effects 
$\H{Hartree}$ and $\H{Layer}$, the more the onset of the magic sector cascade occurs at a larger total filling. 
In particular, strong interactions cause the entire tbg-like nonmagic active band to fill before the magic band fills.
\begin{figure}[h]
  \includegraphics[width=1\textwidth]{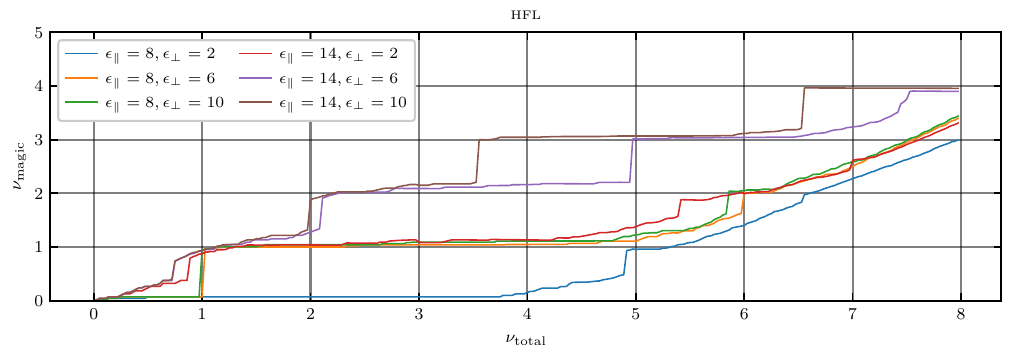}  
  \caption{$\numagic$ as a function of $\nutotal$ at different 
  $\epsilon_\perp$ and $\epsilon_\parallel $ for $N=5$ at strain $\strain = 0.2\%$. }
  \label{app_fig:strainedepsilondependencetpg}
\end{figure}
\subsection{Extended data for large $N$}
\label {app:strainedlargen}
We first examine the effect of changing alternating heterostrain on the data from Figure~\ref{fig:fig_1}d. 
In Fig.~\ref{app_fig:straindepgatecharge}, we compare the charge in the
magic sector for zero and nonzero values of heterostrain at three different gate charges. 
We find a rather weak dependence of the maximal $N$ for $\numagic=4$ on strain, confirming
that the physics at $\numagic = 4$ is mainly governed by electrostatics. On the other hand, when a partial filling
of the magic band occurs, strain dependence is apparent.
\begin{figure}[h]
  \includegraphics[width=1\textwidth]{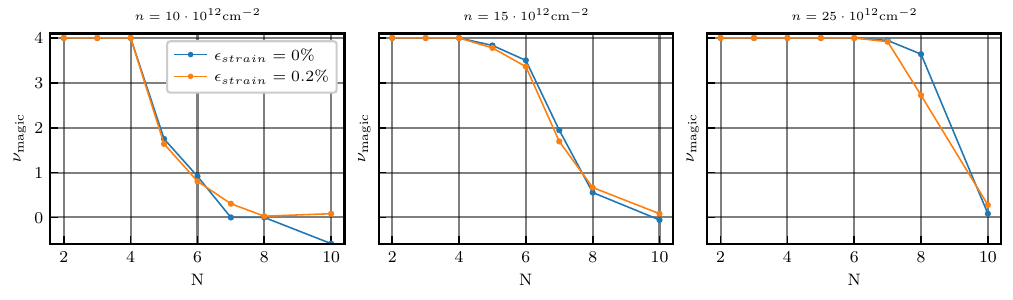}  
  \caption{Strain dependence of charge in the magic 
  sector depending on the gate charge increases from left to right. Here we take
  $\epsilon_\perp=6$,
  $\epsilon_\parallel=10$.}
  \label{app_fig:straindepgatecharge}
\end{figure}

In Fig.~\ref{app_fig:epsilondepgatecharge}, we compare the charge in the magic sector flat bands
for different interaction strengths. We vary $\epsilon_\parallel = 10,14$ and
the ratio $\epsilon_\perp = 6,10$.
The key dependence at lower gate charge $n=10 \cdot 10^{12} \si{cm^{-2}} $ is in fact on $\epsilon_\parallel$, but $\epsilon_\perp$ 
starts to play a role at larger gate densities and large $N$.
\begin{figure}[h]
  \includegraphics[width=1\textwidth]{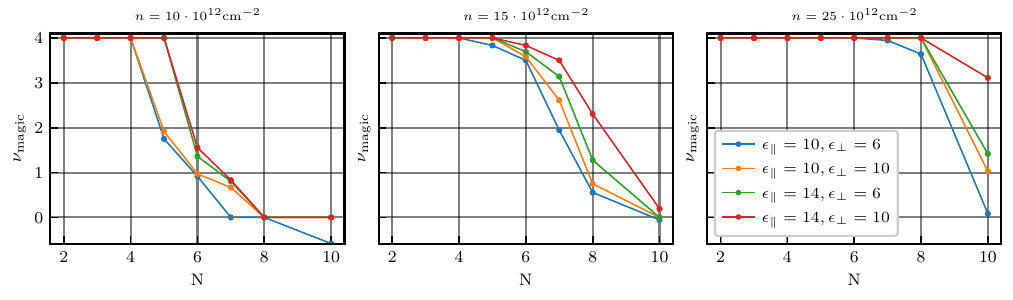}
  \caption{Dependence of the magic 
  sector filling on interaction strength and the number of layers with gate charge $n$ increasing from left to right.
  Here we work at zero strain $\strain = 0.0\%$.}
  \label{app_fig:epsilondepgatecharge}
\end{figure}
\clearpage

We now consider the dependence of our data at full magic sector filling for $\k=1$.
We first consider the parameter dependence of the $\nutotal$ at which the magic bands are fully filled (same as Fig.~\ref{fig:figthree}a).
We consider two different values of strain $\strain = 0, 0.2\%$ and sweep interactions.
As expected, stronger interactions lead to a larger posponement of full magic filling.
\begin{figure}[h]
  \includegraphics[width=1\textwidth]{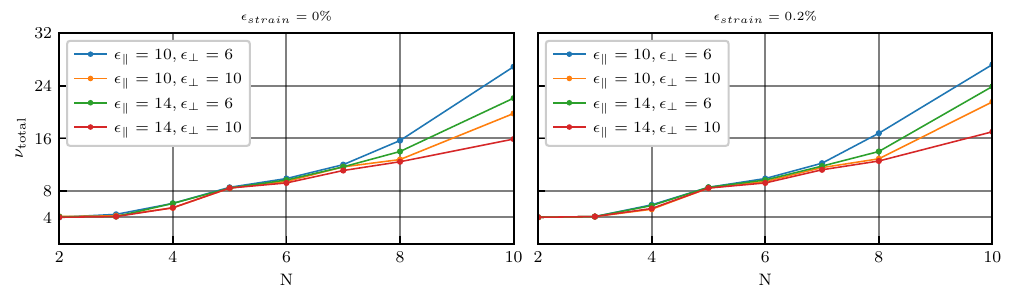}  
  \caption{Interaction strength dependence dependence of total charge needed to fill the magic 
  sector completely.
  Left: $\strain = 0 \%$.
  Right: $\strain=0.2 \%$.}
  \label{app_fig:straindepatmagiccharge}
\end{figure}
In Fig.~\ref{app_fig:epsilondeprs}, we examine the dependence
on interaction parameters of the effective strength of interaction, our $r_s$ data from the main text, Fig.~\ref{fig:figthree}d.
While the unstrained data show relatively little dependence on interaction strength, at finite strain $r_s$ is larger for stronger interactions. Heuristically,
at stronger interactions, the same amount of strain plays a smaller role.
 
\begin{figure}[h]
  \includegraphics[width=1\textwidth]{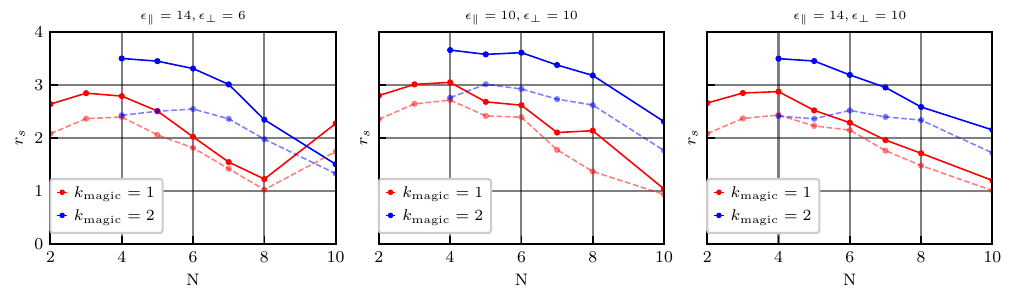}
  \caption{$\epsilon_\parallel$ and $\epsilon_\perp$ dependence of the $r_s$ plot from the main text.
  }
  \label{app_fig:epsilondeprs}
\end{figure}
Lastly, we consider the role of various choices on the $r_s$ plot.
In Figure~\ref{app_fig:rsdeponchoices}, we consider different choices of measuring the bandwidth and $\numagic$.
In particular, in addition to the bandwidth definition from the main text, we could consider the standard deviation of the magic 
band energy distribution $\sigma$ to measure the width of the bands. 
This has the advantage of being less susceptible to outliers than $\mathrm{BW}$ from the main text. For BW, a single $\k$ point at which
there is large mixing can artificially blow up the bandwidth of the band descended from the noninteracting magic band .
Another choice could be not to focus not at $\numagic=4$, but rather at $\numagic =3.6$. 
However, as seen in Fig.~\ref{app_fig:rsdeponchoices}, the advantage of $\k=2$ for $N=5,6$ remains robust to these choices.
\begin{figure}[h]
  \includegraphics[width=1\textwidth]{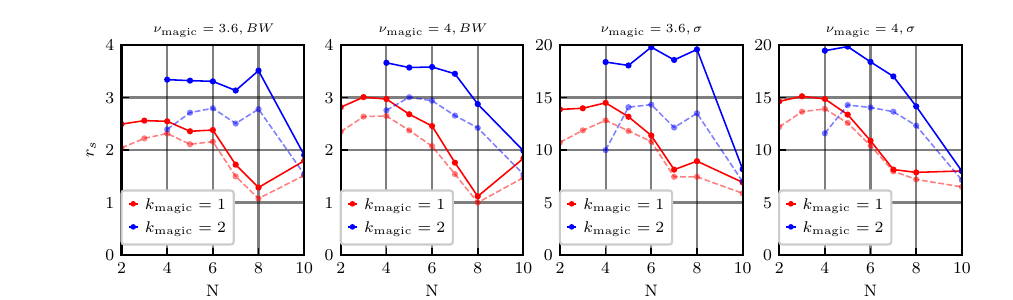}
  \caption{Dependence of the $r_s$ on working at $\numagic=3.6$ or $\numagic=4$ 
  and of using the bandwidth (BW) or standard deviation $\sigma$ as a measure
  of the width of the active magic bands.}
  \label{app_fig:rsdeponchoices}
\end{figure}
\clearpage
\section{Methods}
\label{sec:SM_Methods}
\setcounter{figure}{0}  

\begin{figure}[h]
  \includegraphics[width=1\textwidth]{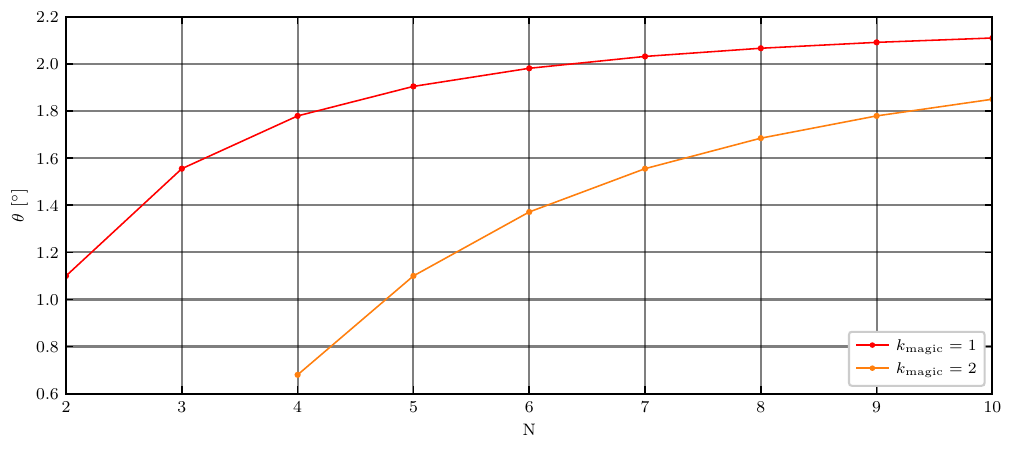}  
  \caption{Physical twist angle for choosing $\k=1$ (red) or $\k =2$ (orange)
  as a function of the layer number $N$.}
  \label{app_fig:twist_angle_ks}
\end{figure}

To obtain the numerical results, we perform self-consistent Hartree-Fock. Our default choice will be a $12\times 12$ $\v k$-space grid.
Our $\v q \neq 0$ interaction is the double-gate screened,
layer dependent (see Sec. \ref{app:layerdepcoulombV}) Coulomb interaction, with gate distance $d_s=40\si{nm}$
and interlayer distance $d_l=0.3 \si{nm}$.
We choose our physical twist angles by the following formula:
    \begin{equation}
    \label{eq:app_physical_angle}
\theta = 2 \cos\left[\frac{\pi \k}{N+1}\right]\,\cdot  1.1^\circ,
    \end{equation}
    chosen so that the effective twist angle of sector $\k$, is the magic
    angle, $\thetaeff{\k}=1.1^\circ$. In Fig.~\ref{app_fig:twist_angle_ks}
 we plot relation
    Eq.~\eqref{eq:app_physical_angle} for the different choices $\k=1,2$ (See
    also Ref.\cite{vishwanathKhalafMagicAngleHierarchy2019} for an equivalent
    plot). 
    This demonstrates that achieving the regime where $k=2$ is in the
    magic regime for $N\ge 5$ is feasible due to the realistic physical twist
    angles of $\theta > 1^\circ$ thus avoiding lattice reconstruction effects.

\subsection{$N \leq 5$}
For the $N\leq 5$ analysis, 
we consider $N_\text{active}=10$ bands and calculate the remote Hartree and Fock contribution 
using $N_{\text{remote}} = 14$ bands below and above the active bands. 
For the heatmap and cascade plots, Figs.~\ref{fig:fig_2}g,h,i, we simulate all four spin/valley flavors, inducing flavor symmetry-breaking by
proposing symmetry-broken trial states at integer fillings. 
For the illustrative band structure and density of states plots, Figs.~\ref{fig:fig_2}d,e,f, 
we use a larger $24\times 24$ grid, but do not include flavor symmetry breaking.
We show the band structures close to $\numagic= 1$.
The cascade and band structure plots are performed at $\epsilon_\parallel = 14$ and $\epsilon_\perp =6$.
\subsection{$N \geq 5$}
For the $N\geq 5$ analysis, 
we consider $N_\text{active}=\max\left[10,2N \right]$ bands and calculate the remote Hartree and Fock contribution 
using $N_{\text{remote}} = \max\left[10, 3N \right]$ bands below and above the active bands. 
This dependence is motivated by the fact that adding a layer adds a band, which we want to include in our analysis, to account for nonmagic sector screening.
We caution, however, that the precise choice is somewhat arbitrary.

For Fig.~\ref{fig:fig_1}a, we work at zero strain and $\epsilon_\parallel=10$, $\epsilon_\perp=6$.
For Figs.~\ref{fig:figthree}a,b, we also work at zero strain and $\epsilon_\parallel=10$, $\epsilon_\perp=6$.
In Figs.~\ref{fig:figthree}c,d we show both zero strain and $\strain=0.2\%$ results.

\subsection{Stability of Hartree-Fock with $\H{layer}$}
In our Hartree-Fock numerics, we ran into an instability for large $\H{layer}$ terms (large filling of large $N$ in combination with a
small out-of-plane constant $\epsilon_\perp$). 
Our system oscillates between states with vertical polarization to 
the top and to the bottom of the sample. Clearly such spontaneously polarized states fail at screening the gate electric field 
and are therefore high energy (see Eq.~\eqref{eq:interacting_hamiltonian}). 
We can understand the appearance of such oscillations by considering mean-field $\H{layer}$ for a state polarized to the top layer in
an infinite density of states system.
In the mean field of such a state, the lowest energy state is the state polarized to the bottom layer.
In this way, there appears an oscillation between opposing vertical polarizations upon iterating Hartree-Fock.
Other terms in Eq.~\eqref{eq:hmfcomplete} make this instability weaker. For example, a finite density of states induces
an energy cost to filling one layer excessively.
We find that explicitly imposing $V_1=V_N=0$ by adding a constant gradient removes this instability, 
at the cost of a slight inaccuracy. 
Numerically, we find that the gradient is small, typically below $\frac{1}{\epsilon_\perp} \si{meV}$.
\end{widetext}
\end{document}